\documentclass[3p,a4paper, 11pt]{elsarticle}

\usepackage[utf8]{inputenc} 
\usepackage[T1]{fontenc}    
\usepackage[hyperref]{xcolor}        
\usepackage[colorlinks=True]{hyperref}       
\usepackage{url}            
\usepackage{booktabs}       
\usepackage{amsfonts}       
\usepackage{amsmath}
\usepackage{amssymb}
\usepackage{nicefrac}       
\usepackage{microtype}      
\usepackage[numbers]{natbib}
\usepackage[noabbrev, nameinlink, capitalise]{cleveref}       

\usepackage{lipsum}         
\usepackage{graphicx}
\usepackage{doi}

\graphicspath{{}}
\usepackage{transparent}
\usepackage{lmodern} 
\usepackage{xspace} 
\usepackage{subcaption}
\usepackage{tikz}
\usepackage{graphicx}
\usepackage{color}
\usepackage{siunitx}
\usepackage{setspace}
\usepackage{lipsum}
\usepackage{bm}             
\usepackage{lineno}         
\usepackage{multirow}       

\input{settings.sty}

\newcommand{\printfontsize}{The current font size is: \the\fontdimen2\font\ }

\title{A PINN Methodology for Temperature Field Reconstruction in the PIV Measurement Plane: Case of Rayleigh-Bénard Convection}

\newbox{\orcid}\sbox{\orcid}{\includegraphics[scale=0.06]{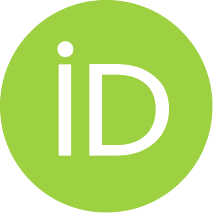}} 
\author[1,2,4]{\href{https://orcid.org/0009-0003-8963-2724}{\usebox{\orcid}\hspace{1mm}Marie-Christine~Volk}\corref{correspondingauthor}}
\author[2,3]{Anne~Sergent}
\author[2]{Didier~Lucor}
\author[1]{Michael~Mommert}
\author[1]{Christian~Bauer}
\author[1,4]{Claus~Wagner}

\cortext[correspondingauthor]{Corresponding author: marie-christine.volk@dlr.de}

\address[1]{Department Ground Vehicles, Institute of Aerodynamics and Flow Technology, German Aerospace Center, D-37073 Göttingen, Germany}
\address[2]{Université Paris-Saclay, CNRS, Laboratoire Interdisciplinaire des Sciences du Numérique (LISN), F-91405 Orsay, France}
\address[3]{Sorbonne Université, Faculté des Sciences et Ingénierie, UFR Ingénierie, F-75005 Paris, France}
\address[4]{Institute of Thermodynamics and Fluid Mechanics, Technische Universität Ilmenau, D-98684 Ilmenau, Germany}

\let\today\relax
\makeatletter
\def\ps@pprintTitle{%
    \let\@oddhead\@empty
    \let\@evenhead\@empty
    \def\@oddfoot{\footnotesize\itshape
         {Preprint — March 2025} \hfill\today}%
    \let\@evenfoot\@oddfoot
    }
\makeatother

\begin{document}
\definecolor{mycolor}{HTML}{1E47C2}

\hypersetup{
  linkcolor=mycolor,       
  urlcolor=mycolor,        
  citecolor=mycolor,       
  filecolor=mycolor,       
}

\newpage

\begin{abstract}
We present a method to infer temperature fields from stereo particle-image velocimetry (PIV) data in turbulent Rayleigh-Bénard convection (RBC)
using Physics-informed neural networks (PINNs). 
The physical setup is a cubic RBC cell with Rayleigh number $\Ra=10^7$ and Prandtl number $\Pr=\num{0.7}$.
With data only available in a vertical plane $A:x=x_0$, the residuals of the governing partial differential equations are minimised in an enclosing 3D domain around $A$ with thickness $\delta_x$.
Dynamic collocation point sampling strategies are used to overcome the lack of 3D labelled information and to optimize the overall convergence of the PINN.
In particular, in the out-of-plane direction $x$, the collocation points are distributed according to a normal distribution, in order to emphasize the region where data is provided.
Along the vertical direction, we leverage meshing information and sample points from a distribution designed based on the grid of a direct numerical simulation (DNS).
This approach points greater attention to critical regions, particularly the areas with high temperature gradients within the thermal boundary layers.
Using planar three-component velocity data from a DNS, we successfully validate the reconstruction of the temperature fields in the PIV plane.
We evaluate the robustness of our method with respect to characteristics of the labelled data used for training: the data time span, the sampling frequency, some noisy data and boundary data omission, aiming to better accommodate the challenges associated with experimental data.
Developing PINNs on controlled simulation data is a crucial step toward their effective deployment on experimental data. The key is to systematically introduce noise, gaps, and uncertainties in simulated data to mimic real-world conditions and ensure robust generalization.
\end{abstract}


\maketitle

\section{Introduction}
Inferring the temperature in thermally-driven flows is highly relevant in fundamental studies and in various technical applications that require precise analysis and control of heat transfer, such as the design of interior ventilation systems in aircraft or train cabins, to optimize air circulation and ensure thermal comfort \cite{dehne_experimental_2024}.
Especially when dealing with such complex technical applications, experiments are crucial to complement, validate or extend theory and simulation.
The flow dynamics can be captured within temporally and spatially highly-resolved three-dimensional (3D) velocity measurements using tomographic particle image velocimetry (tomo-PIV) or particle tracking velocimetry (PTV) \cite{schroder_3d_2023, barta_proptv_2024}.
These methods have also been applied to the canonical experimental setup for thermally-driven flows: Rayleigh-Bénard convection (RBC) (i.e. \cite{schiepel_experimental_2018}).
The simultaneous measurement of spatially resolved velocity and also temperature fields is also feasible using a more elaborated set-up with temperature-sensitive tracer particles (Thermochromic Liquid Crystals, TLCs).
Experiments in RBC were performed for example by \citeauthor{schmeling_development_2010} \cite{schmeling_development_2010}, \citeauthor{schiepel_simultaneous_2021} and \cite{schiepel_simultaneous_2021} or \citeauthor{kaufer_volumetric_2023} \cite{kaufer_volumetric_2023}.
These experimental setups, however, can be challenging, since the TLCs usually have a small color-range, are sensitive towards background light and the colour-to-temperature calibration is not trivial \cite{schmeling_simultaneous_2014}.
An alternative method for obtaining temperature trajectories involves the use of neutrally buoyant, wireless temperature sensors, as demonstrated by \citeauthor{gasteuil_lagrangian_2007} \cite{gasteuil_lagrangian_2007}.
However, the focus of this approach is on obtaining single temperature traces rather than fully resolved spatial temperature fields.
Therefore, the establishment of new methods to reconstruct temperature fields is of great interest.

A significant amount of work has been done in the past on temperature reconstruction from turbulent flows.
One approach for data assimilation or reconstruction in turbulent flows uses methods from Direct Numerical Simulations (DNS) and is called nudging.
It was first systematically applied in \citeauthor{clark_di_leoni_synchronization_2020} \cite{clark_di_leoni_synchronization_2020} to assimilate full-resolved velocity fields out of sparse velocity data.
This technique is also promising for multi-field physical problems, as \citeauthor{bauer_assimilation_2022} \cite{bauer_assimilation_2022} demonstrated that applying it to sparse velocity data from turbulent tomo-PIV measurements allows for temperature field reconstruction.
One challenge in this approach is that errors in the measured velocity fields can significantly impact the evolution and outcome of the temperature field.

Also based on the Navier-Stokes equations, \citeauthor{weiss_temperature_2025} \cite{weiss_temperature_2025} suggested recently a method to solve a Poisson equation for the temperature, using velocity fields of turbulent RBC.
While this technique is computationally efficient, the challenge arises when applying it to experimental data, as it requires spatially and temporally fully resolved fields.

Another method employs machine learning (ML) through the integration of data and partial differential equations (PDEs) within so-called physics-informed neural networks (PINNs).
PINNs are first introduced by \citeauthor{raissi_physics-informed_2019} \cite{raissi_physics-informed_2019} and offer a powerful framework for either forward or inverse problems or assimilation and reconstruction tasks.
An example for forward problems is solving the compressible Euler equations (\citeauthor{wassing_physics-informed_2024} \cite{wassing_physics-informed_2024}).
PINNs are meshfree, and therefore, compared to classic numerical solvers, highly flexible in applications with no access to fully resolved 3D fields.
Several approaches assimilate velocity fields from PTV (e.g. \citeauthor{cai_physics-informed_2024} \cite{cai_physics-informed_2024}, \citeauthor{wang_dense_2022} \cite{wang_dense_2022} and \citeauthor{steinfurth_physics-informed_2024} \cite{steinfurth_physics-informed_2024}) or from different frameworks (i.e. Eulerian or Lagrangian) for efficient pressure recovery in complex moving systems (\citeauthor{sundar_physics-informed_2024} \cite{sundar_physics-informed_2024}).
Recent studies revealed the capabilities of physics-informed neural networks to predict instantaneous temperature fields in turbulent convective flows using 3D velocity fields.
\citeauthor{toscano_inferring_2024} \cite{toscano_inferring_2024} reconstruct temperature fields from PTV data using a physics-informed Kolmogorov-Arnold architecture.
\citeauthor{mommert_periodically_2024} \cite{mommert_periodically_2024} explore the influence of activation functions when reconstructing temperature fields out of fully resolved DNS velocity fields and vice versa and \citeauthor{lucor_simple_2022} \cite{lucor_simple_2022} evaluate the robustness with respect to low resolution in the observation data consisting of sparse temperature data in the bulk and velocity data at the boundaries of the spatial training domain.

Although tomo-PIV and PTV are well-established and extensively validated techniques, they remain demanding in terms of equipment.
Therefore, in this work, we investigate whether it is feasible to rely only on more standard two-dimensional (2D) velocity measurements for temperature reconstruction.
We challenge the PINN to reconstruct temperature fields - in particular with high accuracy in the PIV plane - from stereoscopic velocity data in turbulent RBC by using spatially fully resolved velocity snapshots from DNS (cf. numerical methods in \cref{sec:numerics}).
The use of synthetic data ensures a reliable ground truth, which allows accurate validation of the results.
Our PINN approach employs a padding technique, motivated in \cite{lucor_simple_2022}, by calculating the PDEs in an enclosing 3D domain around the PIV plane and by employing collocation point sampling strategies to further overcome the lack of 3D information (\cref{sec:architecture}).
In order to mimic PIV data and to quantify the challenges of not having access to spatially and temporally fully resolved velocity fields, we test the robustness of the temperature reconstruction with respect to the time span of the data and the time increment between snapshots (\cref{sec:temporal_training_domain}), as well as noisy data and missing labels (\cref{sec:noisy_data}).
In \cref{sec:physical_analysis}, we evaluate the accuracy of the reconstructed temperature fields along with relevant physical quantities.
A key advantage of PINNs is the ability to use automatic differentiation, which provides access to all partial derivatives of the physical fields, including those in the out-of-plane direction.
This allows the calculation and analysis of viscous and thermal dissipation rates (\cref{sec:dissipation}).

\section{Numerical Data Generation} 
\label{sec:numerics}
\begin{figure}[htbp]
 \def\svgwidth{0.4\textwidth}
 \centering
 \begingroup
    \footnotesize  
    \includegraphics{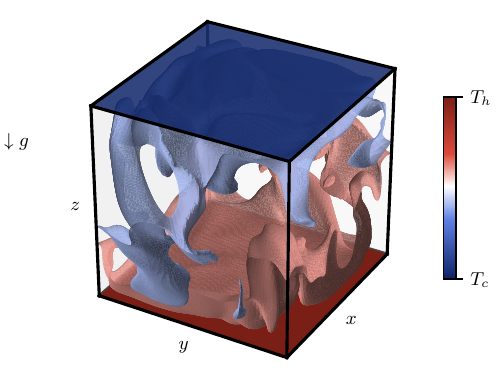}
 \endgroup
 \caption{Schematic of a cubic Rayleigh-Bénard convection cell with height $H$, temperature $T_h$ at the heating plate and $T_c$ at the cooling plate. All boundaries are no-slip and the sidewalls are adiabatic. Temperature isosurfaces at a representative time instant are shown for a DNS with $\Ra=10^7$ and $\Pr=0.7$.}
  \label{fig:rbc_cell}
\end{figure}

\begin{figure}[htbp]
 \centering
 \includegraphics{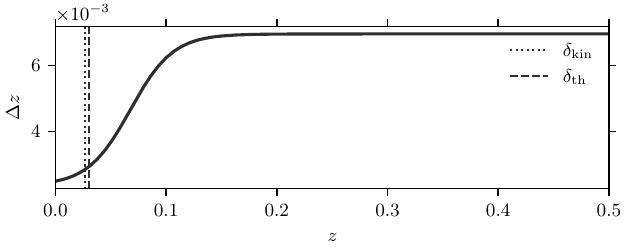}
 \caption{The grid spacing for $\Pr=\num{0.7}$ and $\Ra=10^7$ is progressively refined within the thermal and kinetic boundary layers $\bl_\mathrm{kin/th}$ and uniform in the bulk. The cell height is $H=1$. The required resolution is estimated from \cite{shishkina_boundary_2010} to resolve the local Kolmogorov and Batchelor length scales.}
 \label{fig:plot_grid_spacing}
\end{figure}

We use data generated from a DNS to mimic PIV measurements.
This way, we have access to fully resolved velocity and temperature fields to evaluate the ability of PINNs to reconstruct the temperature field.
A schematic of the geometry and boundary conditions of the cubic RBC cell is depicted in \cref{fig:rbc_cell}, where $T_h$ is the temperature of the heating plate, $T_c$ the temperature of the cooling plate, $g$ the gravitational force in $z$-direction and $H$ the height of the cell.
All boundaries fulfill the no-slip condition and the sidewalls are adiabatic.
The Navier-Stokes equations for an incompressible fluid are the three-dimensional momentum conservation equation together with the mass conservation equation and the energy conservation equation.
In the Boussinesq approximation and in their non-dimensional form, the equations yield
\begin{align}
 \frac{\partial \textbf{u}}{\partial t} + \textbf{u} \cdot \nabla \textbf{u} &= -\nabla p + \sqrt{\nicefrac{\Pr}{\Ra}}\,\nabla^2 \textbf{u} + T \textbf{e}_z\label{eq:navierstokes},\\
 \nabla\cdot \textbf{u}&=0\label{eq:continuity},\\
 \frac{\partial T}{\partial t} + \textbf{u}\cdot \nabla T &= \sqrt{\nicefrac{1}{(\Pr \Ra)}}\,\nabla^2 T\label{eq:heattransfer},
\end{align}
with the three-component velocity vector $\textbf{u}=(u,v,w)$, the temperature $T$, pressure $p$, the Prandtl and Rayleigh number $\Pr=\nicefrac{\tilde\nu}{\tilde\kappa}$, $\Ra=\nicefrac{\tilde g \tilde \alpha \Delta \tilde T \tilde H^3}{\tilde\nu \tilde\kappa}$ and $\textbf{e}_z$ the unit vector pointing upward.
Furthermore, $\tilde\nu$ is the kinematic viscosity, $\tilde\kappa$ the thermal diffusivity and $\tilde\alpha$ the thermal expansion coefficient and the tilde ($\tilde{\cdot}$) denotes the dimensional variables.
The reference velocity used for non-dimensionalization is the free-fall velocity $\tilde u_\mathrm{ref}=\sqrt{\tilde\alpha  \tilde g \Delta \tilde T \tilde H}$, the reference length scale is the cell height $\tilde x_\mathrm{ref}=\tilde H$, the reference time is accordingly $\tilde t_\mathrm{ref}=\nicefrac{\tilde x_\mathrm{ref}}{\tilde u_\mathrm{ref}}$ and the reference pressure is $\tilde p_\mathrm{ref} = \tilde \rho \tilde u^2_\mathrm{ref}$ with the density $\tilde \rho$.
The temperature is non-dimensionalized with $T=\nicefrac{(\tilde T-\tilde T_0)}{\Delta \tilde T}$, where $\tilde T_0=\nicefrac{(\tilde T_h-\tilde T_c)}{2}$ and $\Delta \tilde T =\tilde T_h-\tilde T_c$.
The \cref{eq:continuity,eq:navierstokes,eq:heattransfer} are solved spatially with a finite volume scheme of fourth-order accuracy using a second-order Euler-leapfrog time integration scheme \cite{feldmann_direct_2012}.
In the underlying study, we consider RBC of air with the non-dimensional parameters $\Pr=\num{0.7}$ and $\Ra=10^7$.
To perform a DNS, the mesh widths have to be fine enough to fully resolve the smallest relevant scales of turbulence.
\citeauthor{shishkina_boundary_2010} \cite{shishkina_boundary_2010} derived approximations for suitable grid resolutions for Rayleigh-Bénard convection based on local Kolmogorov and Batchelor length scales.
Based on these estimates, we use the hyperbolic tangent function to construct a grid that progressively refines within the boundaries and transitions to an equidistant spacing in the bulk flow region (cf. \cref{fig:plot_grid_spacing}), resulting in a total number of grid points of $N_{\{x,y,z\}}=\num{162}$, of which 11/12 points are within the $\bl_\text{kin}$/$\bl_\text{th}$ boundary layers in $z$-direction.
The physical fields of the DNS lie on a staggered grid, with the scalar quantities located at the cell centers and the velocity fields positioned at the cell boundaries; therefore, we interpolate the velocity fields to the cell centers using a fourth-order interpolation method.
To mimic data from stereo PIV measurements, the data of the three velocity components on a vertical plane in the center of the cubic domain at $x_0 = \num{0.5}$ is used as labeled data for training.
Unless otherwise stated, we use a time span of $I_t=5$, which corresponds to about half of a large scale circulation (LSC) turnover.
We output every 500th snapshot of the DNS with time step $\Delta t_\text{DNS}=\num{1.5e-4}$, resulting in a dataset consisting of $N_t = 68$ equidistant snapshots.

\section{PINN Configuration}
\label{sec:architecture}
The following section presents the general PINN configuration, including architecture, hyperparameters and optimization, and the methodology for reconstructing temperature fields from planar velocity data.

\subsection{General Architecture and Hyperparameters}
The network consists of a multilayer perceptron (MLP) with $N_\mathrm{L}$ densely connected layers, each containing $N_\mathrm{N}$ sinusoidally activated neurons, except for the output layer, which uses a linear activation.
We use periodically activated neural networks, as they have shown to outperform monotonically activated neural networks in turbulent flow assimilation tasks \cite{mommert_periodically_2024}.
This advantage arises from their resistance to vanishing gradients and their alignment with the inherently periodic nature of turbulent flow dynamics.
The network takes as input a set of 4D spatial and temporal coordinates and outputs the five physical fields.
It can therefore be seen as a function $\bm{f}([x,y,z,t], \bm{\theta})= [u,v,w,T,p]$, where $\bm{\theta}$ is the set of all trainable variables (weights $\bm w$ and bias $\bm b$ parameters).
This architecture leads to the ability to obtain spatial and temporal derivatives of the fields via automatic differentiation \cite{baydin_automatic_2018}.
Using sinusoidal activation functions, the initialization of the weights has a non-negligible impact on the accuracy and convergence speed of the network.
As proposed by \citeauthor{sitzmann_implicit_2020} \cite{sitzmann_implicit_2020}, the weights are randomly initialized by a uniform distribution in the range $\bm w \sim \mathcal{U}\left(-\sqrt{\nicefrac{6}{N_\mathrm{N}^{\mathrm{L}-1}}}, \sqrt{\nicefrac{6}{N_\mathrm{N}^{\mathrm{L}-1}}}\right)$.
All layers are treated equally here, since the introduction of higher frequencies in the first layer, as suggested by \cite{sitzmann_implicit_2020}, does not significantly affect the results (cf. \cite{mommert_periodically_2024}).

\subsection{Loss Functions and Sampling}
\label{sec:loss_functions_sampling}
\begin{figure}[htbp]
 \centering
 \includegraphics{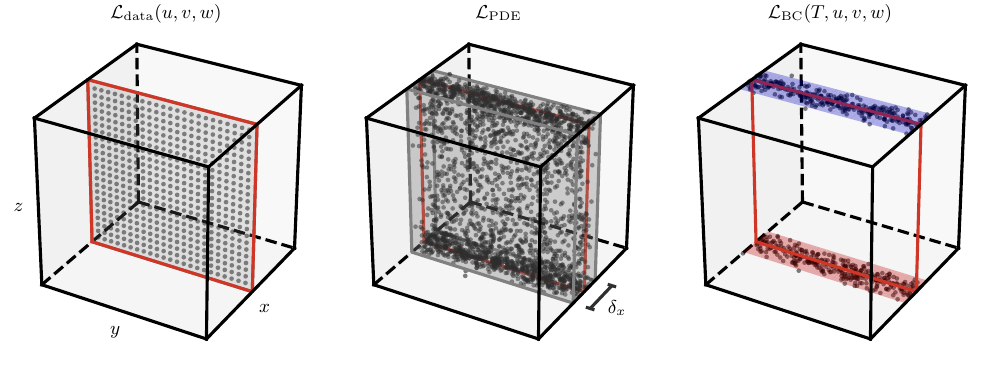}
 \caption{Schematic of the positions of the loss points. Labelled data points for velocity are given only on the vertical plane $A$ (left), whereas the residuals of the PDEs are minimized in a 3D collocation layer with thickness $\delta_x$ (center). Additional Dirichlet data points for velocity and temperature are placed on the top and bottom plates for the boundary condition loss term (right).}
 \label{fig:plot_loss_points}
\end{figure}
PINNs make use of a combined loss function that is empirically evaluated on a set of points, including known data points and selected collocation points where physical information is evaluated.
Our loss function is a weighted sum including data, PDEs and boundary conditions
\begin{align}
 \mathcal{L}= &\lambda_\mathrm{d} \mathcal{L}_\mathrm{data} + \mathcal{L}_\mathrm{PDE} + \mathcal{L}_\mathrm{BC},\\
 \nonumber&\text{with } \mathcal{L}_\mathrm{PDE} = \lambda_\mathrm{m} \mathcal{L}_\mathrm{mom} + \lambda_\mathrm{c} \mathcal{L}_\mathrm{cont} + \lambda_\mathrm{e} \mathcal{L}_\mathrm{energy},\\
 \nonumber&\text{and }\mathcal{L}_\mathrm{BC} =\lambda_\text{t} \mathcal{L}_{\text{BC}(T)} + \lambda_\text{v} \mathcal{L}_{\text{BC}(u,v,w)},
\end{align}
where $\lambda_{[\cdot]}$ are static weights.
An example of visualization of the sampling of the three loss components at a given time is shown in \cref{fig:plot_loss_points}.
The data loss (left) is the mean squared error between the velocity data points and the predictions, using the full spatial DNS resolution on plane $A$
\begin{align}
 \mathcal{L}_\mathrm{data} = \frac{1}{3N_b} \sum_{i=1}^{N_b} \left \| \left( \hat{\textbf{u}}_i-\textbf{u}_i \right)^2\right \| _1, 
\end{align}
where $\hat{\textbf{u}}$ are the true values, $\textbf{u}$ are the reconstructed values and $N_b$ is the batch size.
Since temperature and velocity fields on the top and bottom plates are known, we can augment our dataset by an additional synthetic dataset containing boundary points in regions where no data is given ($z \in \{0,1\}\times x \times y$).
These points are drawn in a domain of thickness $\delta_x$ around $x_0$.
The PDE loss consists of the residuals of \crefrange{eq:navierstokes}{eq:heattransfer}
\begin{align}
 \mathcal{L}_\mathrm{mom} = \frac{1}{3N_b} &\sum_{i=1}^{N_b} \left \| \left( \frac{\partial \textbf{u}_i}{\partial t} + \textbf{u}_i \cdot \nabla \textbf{u}_i +\nabla p_i - \sqrt{\nicefrac{\Pr}{\Ra}}\,\nabla^2 \textbf{u}_i - T_i \textbf{e}_z \right)^2\right \|_1,\\
 \mathcal{L}_\mathrm{cont} = \frac{1}{N_b} &\sum_{i=1}^{N_b} \left( \nabla\cdot \textbf{u}_i \right)^2,\\
 \mathcal{L}_\mathrm{energy} = \frac{1}{N_b} &\sum_{i=1}^{N_b} \left( \frac{\partial T_i}{\partial t} + \textbf{u}_i\cdot \nabla T_i - \sqrt{\nicefrac{1}{(\Pr \Ra)}}\,\nabla^2 T_i \right)^2,
\end{align}
and the corresponding weights are labeled $\lambda_\mathrm{m}$, $\lambda_\mathrm{c}$ and $\lambda_\mathrm{e}$.
The collocation points, where the PDE loss is calculated, are generated in order to help predict the temperature field in the plane $A$ where the labeled velocity data reside.
Previous studies have demonstrated that the performance of PINNs can be improved by strategically placing collocation points in critical regions (e.g., \cite{nguyen_fixed-budget_2023, wu_comprehensive_2023}).
Building on this, we discuss below our sampling strategies to deal with the lack of information in the $x$-direction.

We place the collocation points in a domain around the plane $\mathcal{D} =  [x_0 - \nicefrac{\delta_x}{2}, x_0 + \nicefrac{\delta_x}{2}] \times [0,1] \times [0,1]$ with $\delta_x$ being the thickness (cf. \cref{fig:plot_loss_points} center).
\begin{figure}[htbp]
 \centering
 \begin{subfigure}[t]{0.45\textwidth}
      \includegraphics{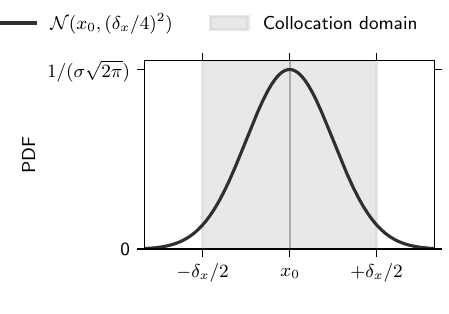}
      \caption{}
      \label{fig:plot_normal_dist}
 \end{subfigure}
 \hfill
 \begin{subfigure}[t]{0.45\textwidth}
      \includegraphics{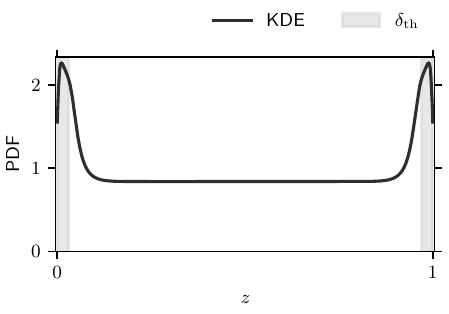}
      \caption{}
      \label{fig:plot_kde_dist}
 \end{subfigure}
 \caption{(a) Collocation points in $x$-direction are sampled using a Normal distribution centered at $\mu=x_0$. The thickness of the collocation layer $\delta_x$ is defined to be the $\pm 2\sigma$-interval of the distribution. (b) The collocation points in the vertical direction are refined in thin layers above and below the heating and cooling plate respectively, by making use of the refinement of the DNS grid shown in \cref{fig:plot_grid_spacing} to resolve the boundary layers, which are characterized by high velocity and temperature gradients.}
\end{figure}
Given the data at $x_0$, the collocation points should be strategically placed to guide the network toward learning the physical solution consistent with the provided data.
The collocation points in $x$-direction are therefore randomly distributed according to a normal distribution with $\mu=x_0$, which is adapted to characteristic physical length scales (see \cref{fig:plot_normal_dist}).
The so-called collocation layer with width $\delta_x$ is defined to be the $\pm2\sigma$-interval of the normal distribution shown in \cref{fig:plot_normal_dist}, thus containing statistically $\SI{95.4}{\percent}$ of the collocation points.
The boundaries are the cell walls at $x=0$ and $x=1$.
This method uses a weighted approach, prioritizing regions with available data while ensuring that all other regions contribute to the accurate reconstruction of the temperature field within $A$.

Since high vertical temperature gradients reside close to the bottom and top walls, where buoyancy increases due to heating, leading to rising hot plumes (bottom plate) and buoyancy decreases, causing sinking cold plumes (top plate), we optimize the positioning of the collocation points in the vertical $z$-direction, to correctly capture these gradients.
Since PINNs generally have difficulty resolving high gradients, a higher density of collocation points is allocated in the boundary layers in comparison to the bulk.
This density is determined by a distribution following the physically motivated distribution of the non-equidistant distribution of the DNS grid points introduced in \cref{sec:numerics}.
We use kernel density estimation (KDE) \cite{parzen_estimation_1962}, where the kernel function is a Gaussian distribution with zero mean and unit variance $K(z)=\mathcal{N}(z,1)$.
The kernel function is evaluated at each of the $N_z$ grid points with a constant kernel bandwidth $h$ and is summed up to form the KDE
\begin{align}
 \text{KDE}(z)=\frac{1}{N_zh}\sum_{i=1}^{N_z}K\left(\frac{z-z_i}{h}\right).
\end{align}
The bandwidth is defined as the mean grid spacing $h=\langle \Delta z \rangle$, resulting in a smooth distribution that is higher near the top and bottom plates and uniform in the bulk (cf. \cref{fig:plot_kde_dist}).
Furthermore, the collocation points are uniformly distributed in time and $y$-direction.

To enhance the network's convergence, the domain is optimally filled by applying the Latin Hypercube Sampling scheme (LHS) \cite{stein_large_1987} together with the aforementioned individual distributions in all four directions.
The use of LHS with a custom distribution ensures that the collocation points follow the desired probability density more effectively, leading to a better representation of the important regions in the domain and thus to a more efficient training of the PINN.

The described method is used to generate a new random set of collocation points at each iteration, where an iteration corresponds to processing a mini-batch of the data.
This effectively covers the physical space and improves generalization by preventing the model from overfitting to a fixed set of points.

\subsection{Training Optimization and Loss Weighting}
\label{sec:optimization}
For an effective training, careful consideration is required when assigning loss weights.
The confidence level of the loss terms is different and we want to follow the most reliable information.
Thus, at this stage, since we are using DNS data for the training, systematic or random measurement errors are not taken into account; consequently, the data loss term is assigned the highest weight $\lambda_\mathrm{d}=\num{1e0}$.
Furthermore, since the velocity data is provided, we weight the Navier-Stokes momentum equation highest from the PDEs ($\lambda_\mathrm{m}= \num{1e-1}$) and the energy equation is weighted less by a factor of ten ($\lambda_\mathrm{e}= \num{1e-2}$).
Since relaxing the constraint of the continuity equation has been shown to improve the training results \cite{lucor_simple_2022}, this term is weighted even less ($\lambda_\mathrm{c}= \num{1e-3}$).
In this particular setup, where the data is given only in a 2D plane, the boundary condition term must be handled carefully to avoid overloading the boundary information which would lead to the trivial solutions of $\textbf{u}=\textbf{0}$ and the heat conduction state $T(z)=0.5-z$.
Consequently, the latter term carries the smallest weight compared to all other loss terms ($\lambda_\mathrm{t}=\lambda_\mathrm{v}= \num{5e-6}$).
Despite the extremely small weights, the boundary conditions are still effectively optimized, as shown by an example training history in \cref{fig:plot_training_history_loss}.

The Adam optimizer with its default configuration is employed for training \cite{kingma_adam_2017}.
A learning rate scheduler monitors the total loss function and gradually reduces the learning rate by a factor of \num{0.8} from an initial value of $\eta_\text{max}=\num{1e-3}$ to a minimum of $\eta_\text{min}=\num{1e-4}$.
In addition, the batch size for both the data and the generation of collocation points is set to $N_\text{b}=4096$.

\section{Evaluation}
Unless otherwise specified, a random spatial and temporal subset of $\SI{5}{\percent}$ of the data points is omitted from the training and used to monitor the training process and for evaluation.
An example of the metrics plotted against the training epochs is provided in \ref{sec:training_history}.
The evaluation metrics used in this study include the mean absolute error (MAE), the coefficient of determination ($\Rtwo$), the relative $\Ltwo$ error, and the relative pointwise error ($\eLtwo$), which is used for plotting
\begin{align}
 \MAE &= \frac{1}{n} \sum_{i=1}^n|\hat y_i- y_i|,\\
 \Rtwo &= 1-\frac{\sum_{i=1}^n (\hat y_i- y_i)^2}{\sum_{i=1}^n (\hat y_i-\overline {\hat y})^2},\\
 \rLtwo &= \frac{\sqrt{\sum_{i=1}^n \left( \hat{y}_{i} - y_{i} \right)^2}}{\sqrt{\sum_{i=1}^n \hat{y}_{i}^2}},\label{eq:rLtwo}\\
 {\eLtwo}_{,i} &= \frac{\left| \hat{y}_{i} - y_{i} \right|}{\sqrt{\sum_{i=1}^n \hat{y}_{i}^2}},
\end{align}
where $\hat y$ are the ground truth values (DNS), $y$ are the reconstructed values and $\overline {\hat y}$ is the mean of the ground truth values.
The default dataset used for training spans a time period of $I_t=5$, which corresponds approximately to half on an LSC turnover time, and is divided into $N_t=68$ equidistant snapshots.
The default PINN configuration consists of $N_\mathrm{L}=10$ densely connected layers with $N_\mathrm{N}=256$ sine activated neurons each, which is the result of a parameter study (see \ref{sec:layer_neuron_variation_study}).
Furthermore, all cases are trained for $10^4$ epochs for comparison.
The final results of the PINN are presented in \cref{sec:physical_analysis}, while the following two sections focus on identifying the critical training parameters.

\subsection{Spatial Training Domain}
\label{sec:spatial_training_domain}
\begin{figure}[htbp]
 \centering
 \includegraphics{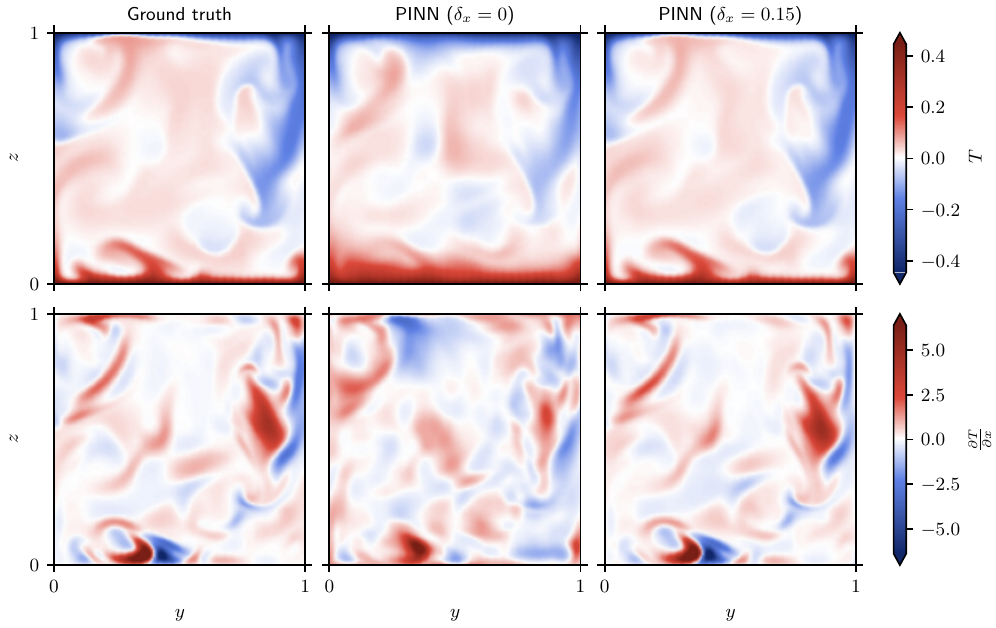}
 \caption{Comparison of planar distribution of temperature fields $T\big|_{x_0}$ (top) and temperature gradient $\nicefrac{\partial T}{\partial x}\big|_{x_0}$ fields (bottom) at a given instant; ground truth data (left), the predictions from PINN configurations with $\delta_x=0$ (center) and $\delta_x=0.15$ (right).}
 \label{fig:plot_dTdx_field}
\end{figure}
Since data and collocation points are only placed in a 2D space, the PINN does not have out-of-plane information and is therefore not capable of accurately predicting the $\nicefrac{\partial}{\partial x}$ derivatives of all physical fields except $\nicefrac{\partial u}{\partial x}$, which can be obtained from the continuity equation (\cref{eq:continuity}).
Therefore, for the reconstruction of the in-plane temperature field, it is crucial to determine an appropriate collocation layer thickness $\delta_x$.
\Cref{fig:plot_dTdx_field} shows a comparison of planar temperature fields $T\big|_{x_0}$ and temperature gradient fields $\nicefrac{\partial T}{\partial x}\big|_{x_0}$ from the ground truth data (left) and two different PINN reconstructions with $\delta_x=0$ (middle) and $\delta_x=0.15$ (right), i.e. a 2D and 3D collocation domain.
It can be seen that the PINN trained with $\delta_x=0$ is not able to predict the structures of the ground truth $\nicefrac{\partial T}{\partial x}$ field.
Some key structures in the temperature field are still recognizable, but the overall field is significantly distorted compared to the ground truth; whereas for $\delta_x=0.15$ the field closely resembles the ground truth field.
\begin{figure}[htbp]
 \centering
 \includegraphics{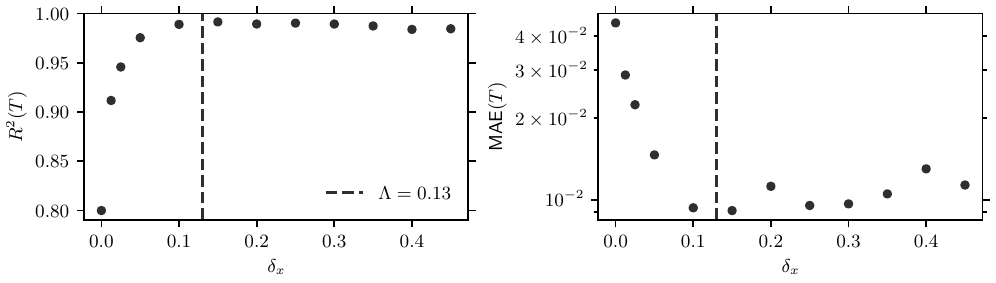}
 \caption{Coefficient of determination $\Rtwo$ (left) and MAE values (right) of the temperature reconstruction vs. the thickness $\delta_x$ of the collocation layer. The lowest-performing result for the temperature prediction is obtained using only a 2D collocation domain ($\delta_x=0$). The results reach a plateau with increasing $\delta_x$, as training into regions too distant from the data points provides no benefit.}
 \label{fig:plot_delta_var}
\end{figure}
\Cref{fig:plot_delta_var} shows $\Rtwo$ and $\MAE$ for the predicted temperature field in dependence of the thickness $\delta_x$.
The training time increases for larger training domains because the spatial collocation point density remains the same for each case.
Increasing $\delta_x$ leads to increasing $\Rtwo$ scores and decreasing MAE values until $\delta_x\approx0.15$.
Further expanding the collocation layer does not improve the results.
For the largest tested values $\delta_x\ge0.4$, there is even a decrease in the $\Rtwo$ score and increase in the $\MAE$ values.
In regions far from $x_0$, where no data is available, the PINN generates its own physically meaningful field.
However, this field may not fully correspond to the ground truth in these regions, potentially introducing an unwanted influence on the solution at $x_0$.
Therefore, a good choice for the thickness of the training domain is the length scale where the velocity field is still statistically correlated with the field itself at $x_0$.
Following \citeauthor{thacker_comparison_2010} \cite{thacker_comparison_2010}, we compute the horizontal integral length scales of the velocity components averaged over $\num{73}$ 3D snapshots equidistantly distributed over a large time span of $t = 500$.
The average of all components in both horizontal directions is $\Lambda=0.13$, which agrees well with the optimal collocation layer thickness $\delta_x=0.15$.

\subsection{Temporal Training Domain}
\label{sec:temporal_training_domain}
Since, unlike other studies, we do not specify initial conditions for the training, the result of the training depends on the time span of the available data.
Therefore, in this section, we analyze how the time span and the temporal resolution of the given data snapshots influence the accuracy of the reconstructed temperature field.

The PINN presumably needs to be trained over a sufficient time span to learn the dynamics of the flow and also the $\nicefrac{\partial}{\partial t}$ derivatives of the fields, analogous to the spatial domain requirement discussed in the previous \cref{sec:spatial_training_domain}.
\begin{figure}[htbp]
 \centering
 \includegraphics{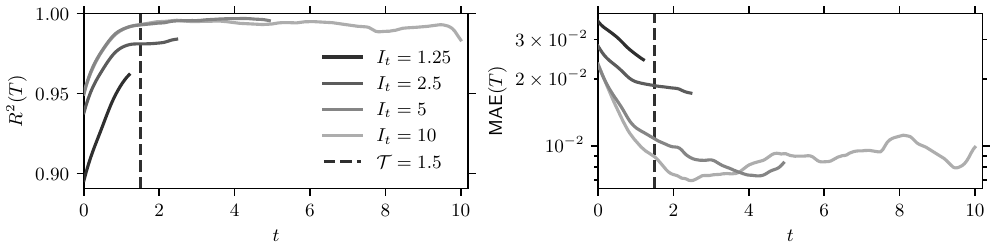}
 \caption{Comparison of $\Rtwo$ (left) and MAE (right) values of temperature reconstruction for PINNs trained over different time spans, with a snapshot interval of $\Delta t=0.06$.}
 \label{fig:plot_R2_time_tVar}
\end{figure}
We trained the PINN with different time spans of data $I_t\in \{1.25, 2.5, 5, 10\}$.
The longest run covers approximately a full turnover of the LSC.
\Cref{fig:plot_R2_time_tVar} shows the $\Rtwo$ and $\MAE$ metrics for the temperature reconstruction with data sampled at time intervals of $\Delta t=0.06$.
For a fair comparison, the runs have processed the same number of collocation points per time unit during their training.
The evaluation metrics are computed using PINN predictions on plane $A$ (full resolution).
It can be observed that, in all cases, the results are worst at the beginning of the time span.
Only the two longest runs are long enough for $\Rtwo$ to reach a plateau, while towards the end of the time span, the results slightly deteriorate again (cf. $I_t=10$).
The best result for a single snapshot is $\Rtwo=0.997$ and is obtained with $I_t = 5$ at $t=3.90$.
The results are poorest at the boundaries, but particularly at early time instances, indicating that earlier snapshots aid the PINN in reconstructing subsequent fields, but this effect is not as strong the other way round.
From this we conclude, that the performance of the PINN depends on the direction of the time.
This phenomenon is related to the irreversible nature of turbulent flows, driven by dissipation.
For comparison, we compute the integral time scale from the auto-correlation of time series of all velocity components.
Averaging over a total of 648 signals, each with a duration of $I_t=30$, distributed over 27 positions in the cell, we obtain $\mathcal{T}=1.5$.
As visible in \cref{fig:plot_R2_time_tVar}, the PINN requires approximately an integral time scale at the beginning of the temporal training domain until the $\Rtwo$ result reaches a plateau, which is an analogous observation to the minimum spatial domain size requirement in \cref{sec:spatial_training_domain}, which was found to be one integral length scale of the flow.
This suggests that, in the absence of initial conditions, the PINN requires a characteristic time scale to constrain its degrees of freedom and accurately reconstruct the physical fields.
However, to determine whether a general principle holds, it would be necessary to test RBC cases with varying parameters.

\begin{figure}[htbp]
 \centering
 \includegraphics{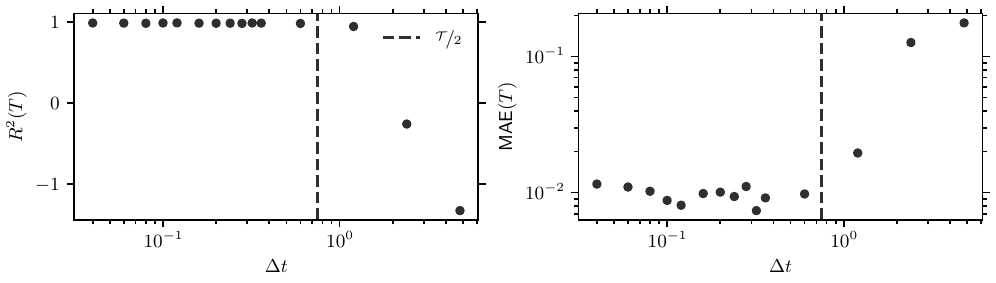}
 \caption{Variation of the time increment $\Delta t$ between the snapshots using a time span $I_t=5$.}
 \label{fig:plot_dt_var}
\end{figure}
Since PIV experiments cannot achieve the temporal resolution of a DNS, we also investigate the influence of the data snapshot acquisition frequency (here reported as the time increment between two consecutive snapshots).
\Cref{fig:plot_dt_var} shows the $\Rtwo$ and $\MAE$ metrics of the temperature reconstruction using a time span $I_t=5$ and different time increments $\Delta t \in [0.04, 4.8]$.
We observe that $\Rtwo(T)$ and $\MAE(T)$ remain consistent as $\Delta t$ increases over a wide range of values.
However, for a step size larger than half the integral time scale, both metric results collapse.
In other words, the time span of an integral time scale must be sampled at least at twice the frequency, to be able to reconstruct the time evolution of the flow structure with the PINN.
This is consistent with the Nyquist-Shannon sampling theorem \cite{shannon_communication_1949}, meaning that the intermediate time scales, such as the integral time scale, must be resolved to allow the PINN to reconstruct the temporal evolution in between.
This is an interesting result that is probably not specific to our setup, but rather applicable to any reconstruction tasks involving time-dependent flow structures using PINNs.

\begin{figure}[htbp]
 \centering
 \includegraphics{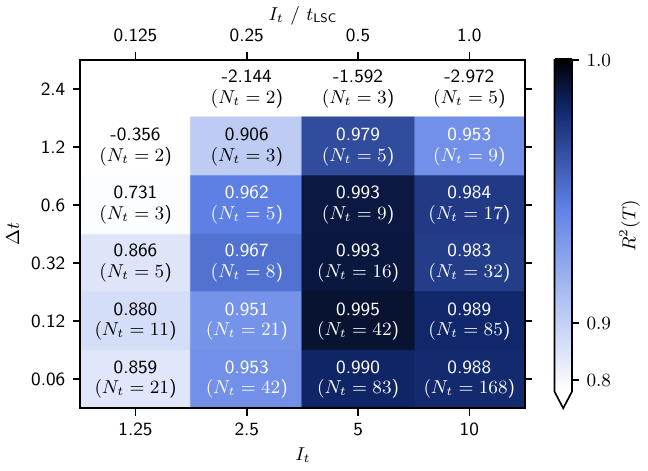}
 \caption{Variation of the time span $I_t$ and the time increment $\Delta t$ between the snapshots. $N_t$ is the number of data snapshots used. The average time of an LSC turnover is approximately $t_\text{LSC}\approx10$. The plot has a power-normal color scale with a gamma value of 15, which preserves details in high-scoring areas.}
 \label{fig:plot_timespan_dt_matrix}
\end{figure}
An overview of the results of the parametric study with different time spans and time step sizes is shown in \cref{fig:plot_timespan_dt_matrix}.
The $\Rtwo$ metric is computed using the full DNS resolution of snapshots with $\Delta t=0.06$ over the entire temporal training domain ($t>\mathcal{T}$, for cases with sufficiently long time spans).
Regardless of $\Delta t$, it can be seen that the time spans $I_t=\numlist{1.25;2.5}$ are not long enough and therefore produce the worst results.
The best results are obtained for $I_t=5$, which corresponds to about half of an LSC turnover time, with no significant difference when using either $N_t=\numlist[list-final-separator = \text{ or }]{9;16;42}$ snapshots.

\subsection{Physical Analysis of the PINN-Reconstructed Fields}
\label{sec:physical_analysis}
In this section, we analyze the physical properties of the reconstructed temperature fields.
All results presented below are based on a temporal training domain of $I_t=5$ with a snapshot interval of $\Delta t =0.075$.
For the statistical analysis, twelve cases are considered, each starting at a different time point of the simulation.
As observed in the previous \cref{sec:temporal_training_domain}, the results within the range of one integral time scale at the beginning of the temporal training domain are the worst.
Therefore, we exclude this range from the evaluation.

\subsubsection{Temperature Reconstruction}
\begin{figure}[htbp]
 \centering
 \includegraphics{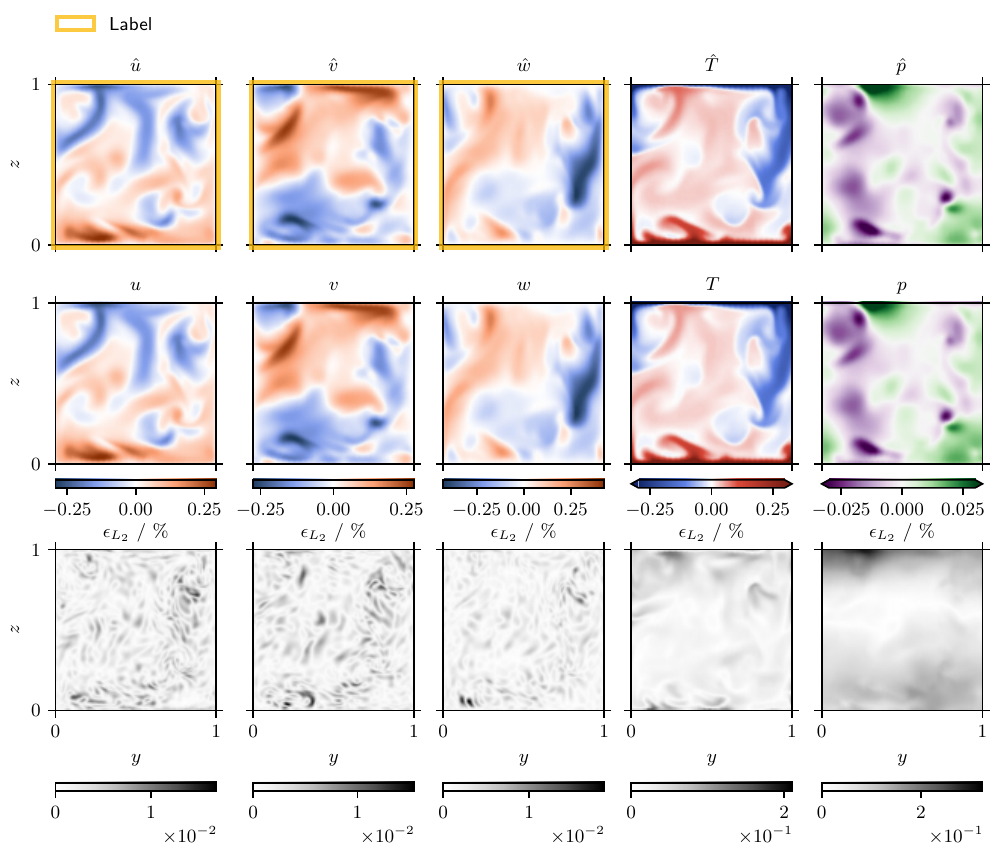}
 \caption{Comparison of ground truth ($\hat\cdot$, top row) and reconstructed fields (center row) on plane $A$ and their relative pointwise error field $\eLtwo$ (bottom row) at a time instant $t=\nicefrac{I_t}{2}$.}
 \label{fig:plot_prediction}
\end{figure}
\Cref{fig:plot_prediction} displays the ground truth fields (top row), the reconstructed fields (center row) and the relative pointwise error fields $\eLtwo$ (bottom row) of all five physical quantities on plane $A$ in the middle of the temporal training domain ($t=\nicefrac{I_t}{2}$).
The absolute pressure values cannot be determined because the PDE (\cref{eq:navierstokes}) contains only the pressure gradient $\nabla p$.
Therefore, for better comparison, the respective field averages are subtracted from the pressure ground truth and the reconstructed field.
Since velocity data is provided for training, the network successfully resolves all visible structures, which is also reflected in the small-scale error fields (\cref{fig:plot_prediction}, bottom).
There is no remarkable difference in the $\eLtwo$ fields of the out-of-plane component $u$ and the in-plane horizontal component $v$.
Furthermore, the visible structures of the temperature and pressure fields are reconstructed in all details by the PINN (\cref{fig:plot_prediction}, middle row).
All plumes, corner eddies and the boundary layers are successfully reconstructed.
However, for the reconstructed temperature and pressure fields, the error field scales are larger and partially coincide with high gradient regions of the true field.
This indicates that the unknown quantities inferred by the PINN have the highest uncertainties, particularly in regions with steep gradients.

\begin{figure}[htbp]
 \centering
 \includegraphics{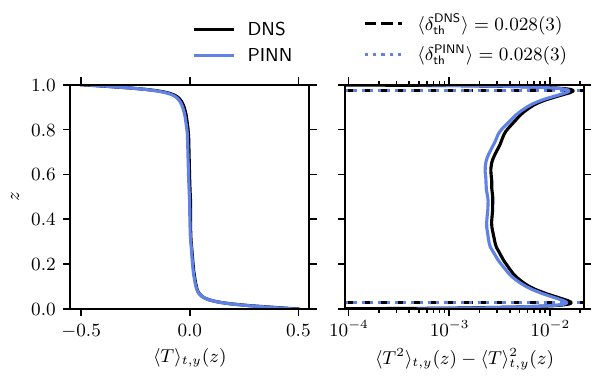}
 \caption{The temperature field in plane $A$, averaged over $t$ and $y$ (left), along with its variance (right) as a function of the height $z$ from DNS and PINN reconstructions over a total time span of $I_t=42$.}
 \label{fig:plot_T_profile}
\end{figure}
To further investigate the reconstruction of thermal boundary layers, \cref{fig:plot_T_profile} shows the temperature profile, averaged over $t$ and $y$, as a function of the height $\langle T\rangle_{t,y}(z)$ and its variance $\langle T^2\rangle_{t,y}(z)-\langle T\rangle_{t,y}^2(z)$, computed with true and reconstructed fields on plane $A$ from a total time span of $I_t^\text{tot} = 42$.
The resulting temperature profile has the characteristic shape of the RBC, with the reconstructed mean temperature profile closely matching the ground truth.
Slight differences are noticeable in the variance profile within the bulk of the cell.
From the peaks of the variance, we can determine the thickness of the thermal boundary layer.
Averaging over the upper and lower layer thickness, we obtain the same value of $\bl_\text{th}^\text{DNS} = \bl_\text{th}^\text{PINN} = \num{0.028\pm0.003}$.
Using the Nusselt number $\Nu=16.2$ derived from the entire cubic cell in \cite{kaczorowski_turbulent_2014}, we obtain the boundary layer thickness $\bl_\text{th}=\nicefrac{1}{(2\Nu)}=0.031$, which is consistent with the values we obtain in \cref{fig:plot_T_profile}.

\begin{figure}[htbp]
 \centering
 \includegraphics{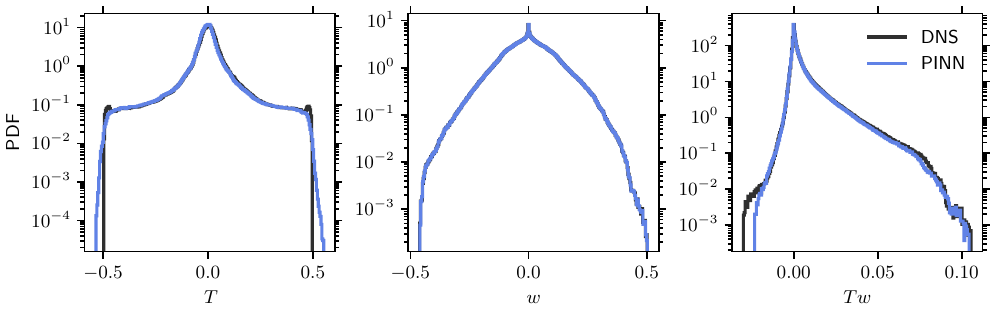}
 \caption{Probability density functions of the true and reconstructed fields of temperature (left), vertical velocity component (center) and vertical convective heat flux (right) over a total time span of $I_t=42$.}
 \label{fig:plot_Tw_histograms}
\end{figure}
To examine the reconstruction of the temperature in more detail, we plot the probability density functions (PDF) of the temperature (left), the vertical velocity component $w$ (middle), and the vertical convective heat flux $Tw$ (right) in \cref{fig:plot_Tw_histograms}.
The reconstructed temperature is generally consistent with the DNS histogram (\cref{fig:plot_Tw_histograms}, left); however, for the extreme values, instead of showing a peak, the values $T_h=0.5$ and $T_c=-0.5$ are exceeded.
This is possible due to the linear activation in the last layer.
While no deviations are observed for $w$ in \cref{fig:plot_Tw_histograms} (middle), there is generally good agreement between the DNS and reconstructed fields in the $Tw$ histogram (\cref{fig:plot_Tw_histograms}, right).
However, in the left tail, the PINN histogram exhibits a steeper slope and an earlier cut-off.
This indicates difficulties in resolving regions where positive temperature is not associated with positive velocity, and vice versa.

\begin{figure}[htbp]
 \centering
 \includegraphics{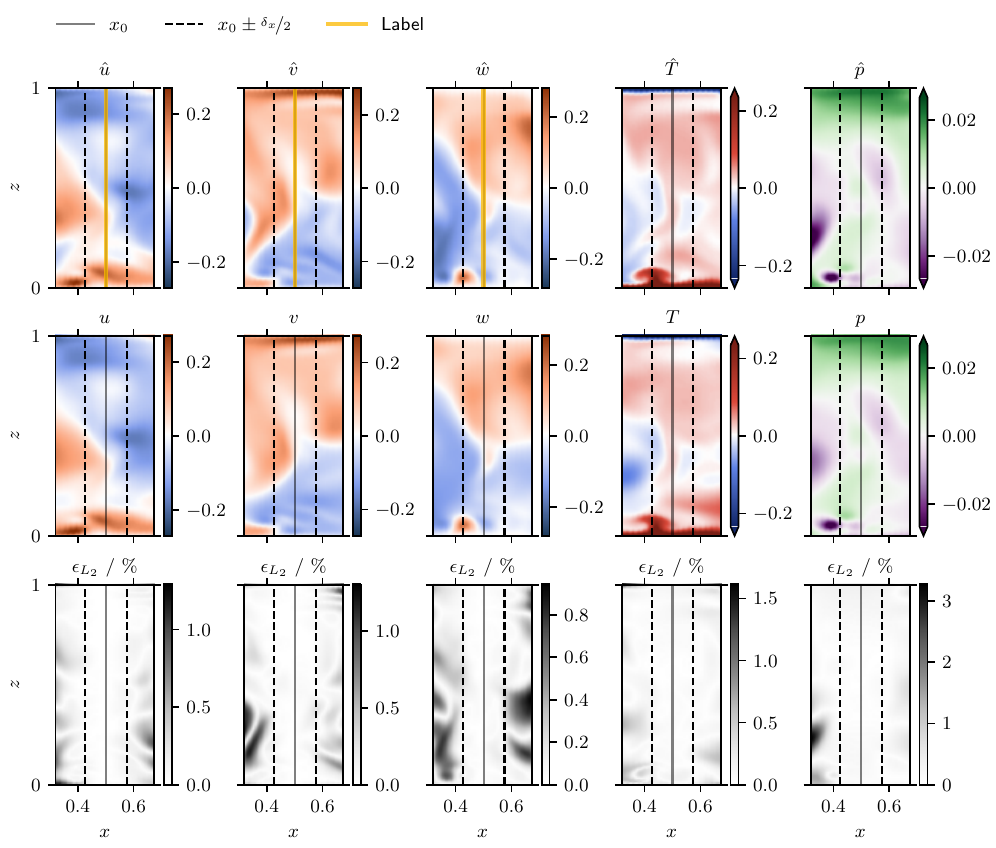}
 \caption{Comparison of DNS (top row) and reconstructed out-of-plane fields (middle row) and their relative pointwise error fields $\eLtwo$ (bottom) at $y=0.5$ and $t=\nicefrac{I_t}{2}$. Velocity data is given at $x_0$ and the width of the training domain is $\delta_x=0.15$.}
 \label{fig:plot_prediction_offplane}
\end{figure}
Since the training is performed in a 3D domain, we can also examine the reconstructions out-of-plane.
\Cref{fig:plot_prediction_offplane} shows a section of the vertical plane at $y=0.5$.
At this view angle, data for training is given only at the vertical center line ($x=x_0$) and training is accomplished in the area $]x_0-\nicefrac{\delta_x}{2},x_0+\nicefrac{\delta_x}{2}[$ marked by the dashed vertical lines.
The top row shows the ground truth fields, followed by the reconstructed fields in the middle row, and the relative pointwise error fields $\eLtwo$ at the bottom.
It is noteworthy, that the PINN fully reconstructs the out-of-plane velocity fields within the $\delta_x$ domain.
Even outside of the domain, where \SI{5}{\percent} of the collocation points are placed, the fields highly resemble the ground truth.
This is clearly visible in the $u$ field at $x\approx0.4$ near the bottom (\cref{fig:plot_prediction_offplane}, far left column), where the PINN fully reconstructs a rightward-streaming region.
Since only a small fraction of the collocation points are placed outside the collocation layer and the distance to the given data increases, the velocity error fields exhibit the highest relative errors at the left and right edges of the image.
Furthermore, also the structures of the temperature and pressure fields are completely captured inside the collocation layer.
Notable is the reconstruction of a hot plume at $x\approx0.3$, which then blurs to the left outside the collocation domain.
The same location in the pressure field shows a low pressure region, which is also successfully captured by the PINN.

\subsubsection{Viscous and Thermal Dissipation Rates}
\label{sec:dissipation}
\begin{figure}[htbp]
 \centering
 \includegraphics{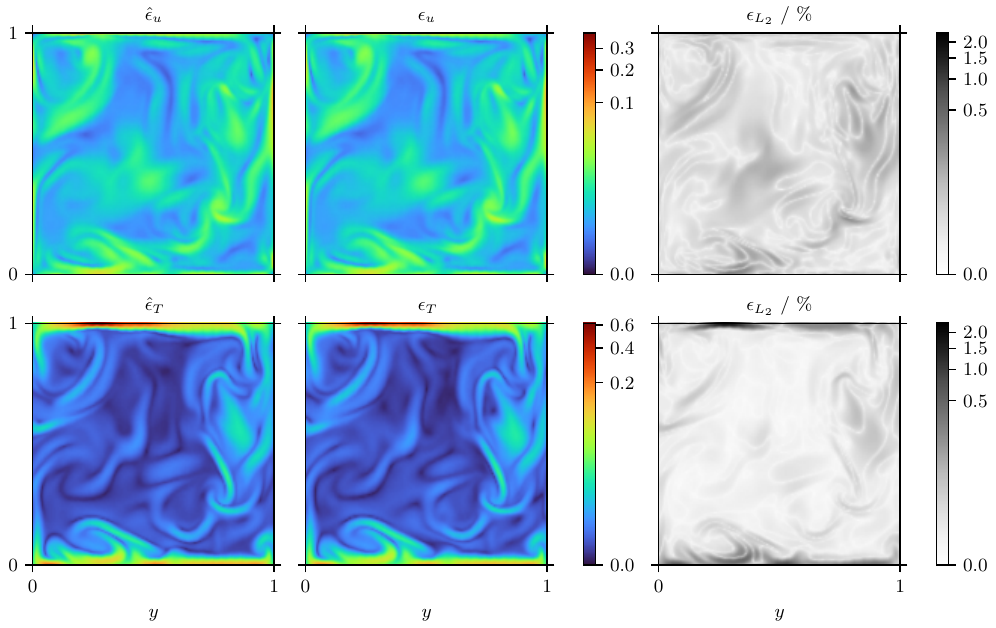}
 \caption{Viscous (top row) and thermal (bottom row) dissipation rates at plane $A$ at a given time instant ($t=\nicefrac{I_t}{2}$). The ground truth fields ($\hat\cdot$) are shown on the left, the PINN predicted fields obtained by automatic differentiation are shown in the middle, and the relative error fields are shown on the right.}
 \label{fig:plot_dissipation}
\end{figure}
A major advantage of using a PINN is that the automatic differentiation provides access to all partial derivatives of the five physical fields.
This includes derivatives in the out-of-plane direction, even though the data is given in a plane only.
As a result, we are able to analyze the viscous ($\epsilon_u$) and thermal ($\epsilon_T$) dissipation rates, which can be derived from \cref{eq:continuity,eq:navierstokes,eq:heattransfer} (see e.g. \cite{shraiman_heat_1990}), and which are
\begin{align}
 \epsilon_u &= \sqrt{\frac{\Pr}{\Ra}}\sum_{i,j}\left(\frac{\partial u_i}{\partial x_j}\right)^2,\label{eq:viscous_diss}\\
 \epsilon_T &= \frac{1}{\sqrt{\Ra \Pr}} \sum_i \left(\frac{\partial T}{\partial x_i}\right)^2.\label{eq:thermal_diss}
\end{align}
\Cref{fig:plot_dissipation} shows the viscous (top row) and thermal (bottom row) dissipation rates in plane $A$ at a given time instant.
The left column shows the ground truth fields ($\hat \cdot$) obtained by central differences from the DNS fields, the middle column shows the dissipation fields obtained by automatic differentiation from the PINN, and the right column shows the relative pointwise error fields ($\eLtwo$).
According to the scaling theory proposed by \citeauthor{grossmann_scaling_2000} \cite{grossmann_scaling_2000,stevens_unifying_2013}, four different main scaling regimes are obtained, depending on the boundary layer and bulk contributions of the global dissipation rates.
Our parameters $Pr=0.7$ and $Ra=10^7$ are assigned to the regime, where $\epsilon_u$ is dominated by the bulk and $\epsilon_T$ is dominated by the boundary layers, which is reflected in \cref{fig:plot_dissipation}.
The structures in both reconstructed (viscous and thermal) dissipation fields (\Cref{fig:plot_dissipation}, middle column) closely resemble the ground truth (\Cref{fig:plot_dissipation}, left column).
The error plots (\Cref{fig:plot_dissipation}, right column) give a better insight into the location of the uncertainties.
For the viscous dissipation, higher error regions are distributed throughout the cell, while for the thermal dissipation, the highest error values are obtained near the top and bottom walls, where the highest dissipation rates are found.
Since velocity data is provided for training, we expect the overall error for viscous dissipation to be smaller than for thermal dissipation.
The relative $\Ltwo$ errors computed with data on plane $A$ from the time instant shown in \cref{fig:plot_dissipation} are $\rLtwo(\epsilon_u)=\SI{9.3}{\percent}$ for the viscous dissipation, and $\rLtwo(\epsilon_T)=\SI{19.7}{\percent}$ for the thermal dissipation.

\subsection{Robustness Towards Noisy Data and Missing Labels}
\label{sec:noisy_data}
\begin{figure}[htbp]
 \centering
 \includegraphics{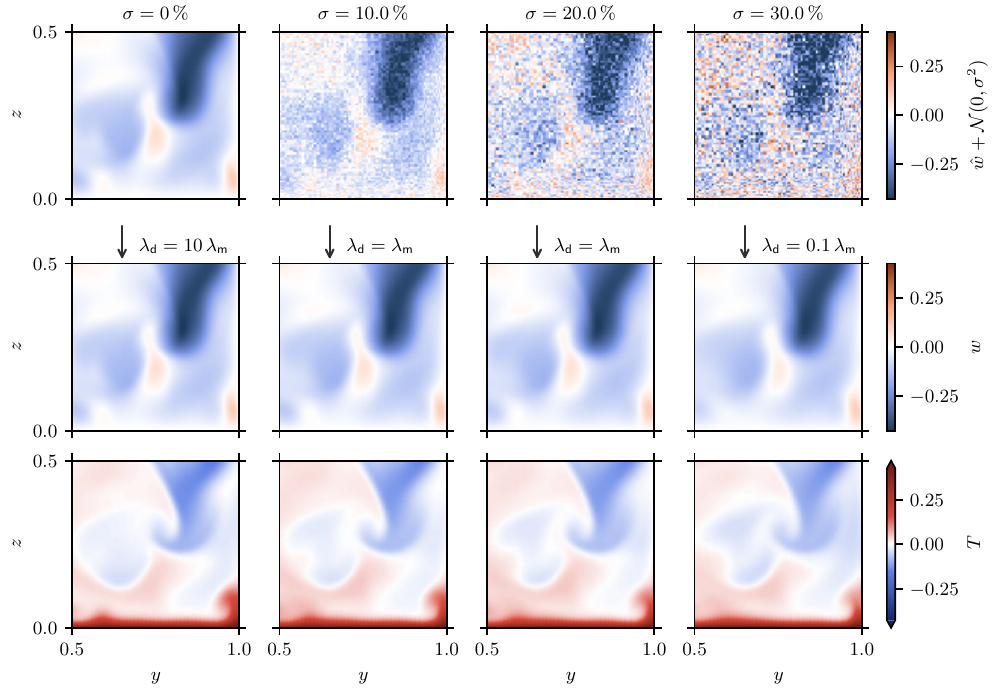}
 \caption{Ground-truth vertical velocity fields in a quadrant of plane $A$ with varying intensities of Gaussian noise $\mathcal{N}(0,\sigma^2)$ relative to the maximum amplitude of the velocity component (top row), reconstructed vertical velocity fields (middle row) and reconstructed temperature fields (bottom row). The weight of the data loss term ($\lambda_\mathrm{d}$) is reduced compared to the weights of the PDE loss terms for higher $\sigma$.}
 \label{fig:plot_noisy_DNS_pred_fields}
\end{figure}
\begin{figure}[htbp]
 \centering
 \includegraphics{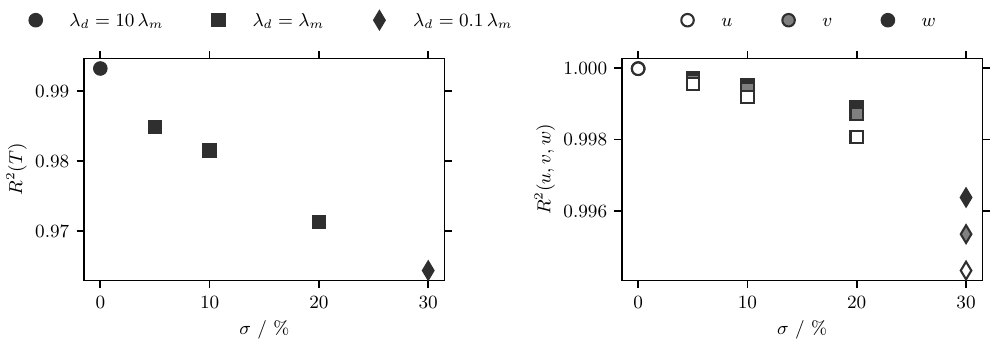}
 \caption{Comparison of temperature (left) and velocity (right) reconstruction results at different intensities of Gaussian noise $\mathcal{N}(0, \sigma^2)$, expressed as a percentage of the maximum velocity component amplitude. At higher noise ratios, the weighting of the data loss term $\lambda_\mathrm{d}$ has to be decreased compared to the PDE loss terms.}
 \label{fig:plot_noise_var_T_vel}
\end{figure}
To assess the potential applicability of our method to more realistic PIV data, we evaluate the robustness of the model to noisy training data.
We superimpose Gaussian white noise $\mathcal{N}(0,\sigma^2)$ on the velocity fields with standard deviations $\sigma$ of different levels:  \SIlist{0; 10; 20; 30}{\percent} relative to the maximum amplitude of each velocity component.
Generally, in stereo PIV measurements, the out-of-plane velocity component is the one more sensitive to measurement uncertainty; here, for the sake of simplicity, all velocity components are treated equally.
So far, we have only used DNS data, which is why the data loss term was weighted the highest in the PINN.
However, with error-prone data, the weight of the data loss term ($\lambda_\mathrm{d}$) must be reduced relative to the PDE loss weights, to allow the PINN more flexibility in reconstructing the velocity fields.
We tested weights in increments of powers of 10.
\Cref{fig:plot_noisy_DNS_pred_fields} shows the vertical velocity field with different levels of Gaussian noise (top), the reconstructed velocity field (center) and the reconstructed temperature field (bottom).
The relationship between the weight of the data loss term ($\lambda_\mathrm{d}$) and the weight of the momentum conservation equation ($\lambda_\text{m}$) is shown, since the balance between the individual PDE loss weights remains unchanged (cf. \cref{sec:optimization}).
To better visualize the differences, only one quadrant of plane $A$ is shown.
The assimilated velocity fields are very similar to the ground truth data (\cref{fig:plot_noisy_DNS_pred_fields}, second row), although there is a slight loss of amplitude at higher $\sigma$.
This effect is even more pronounced in the temperature field (\cref{fig:plot_noisy_DNS_pred_fields}, bottom).
While the overall structures are generally well captured, their intensity decreases slightly at higher $\sigma$, as seen, for example, in the corner plume in the lower right of the temperature fields (\cref{fig:plot_noisy_DNS_pred_fields}, bottom).
\Cref{fig:plot_noise_var_T_vel} shows the corresponding $\Rtwo$ values for the temperature reconstruction (left) and the velocity reconstruction (right).
The $\Rtwo$ value of the temperature reconstruction decreases with increasing noise ratio from $\Rtwo(T, \SI{0}{\percent})>0.99$ to $\Rtwo(T, \SI{30}{\percent})\approx0.96$.
The same applies for the velocity; however, in this case, we still consistently obtain values of $\Rtwo>0.99$.
When comparing the different velocity components, it is noticeable that the result for $u$ is the worst throughout.

\begin{figure}[htbp]
 \centering
 \includegraphics{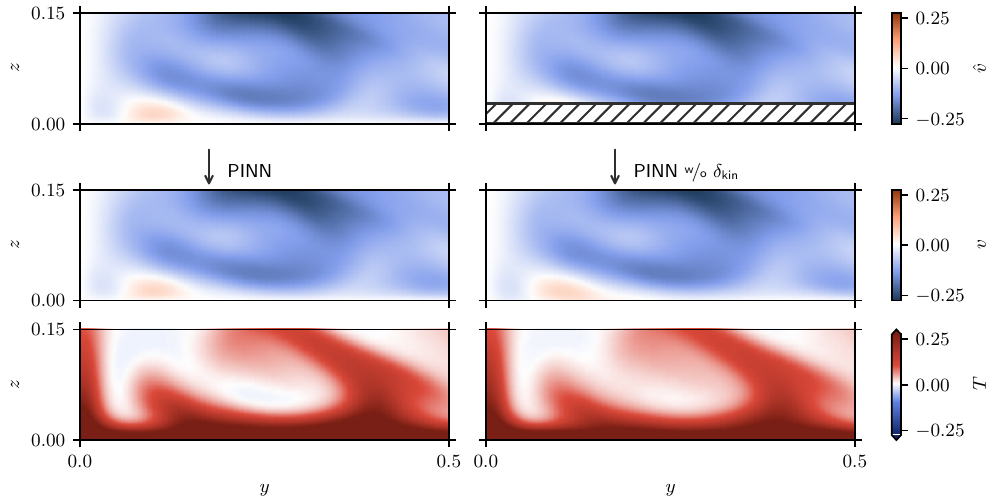}
 \caption{Comparison of the PINN trained with the spatial DNS resolution on slice $A$ (left) and when omitting the velocity data in the kinematic boundary layer (right). The ground truth horizontal velocity field $\hat v$ is shown in the top row, the reconstructed velocity field $v$ in the middle row and the reconstructed temperature field in the bottom row. In order to focus on the boundary layers, only a subset of $A$ is shown.}
 \label{fig:plot_omitBL_fields}
\end{figure}
While in DNS the mesh resolution is usually refined in the boundary layers, in experimental setups it is difficult for the tracer particles to reach the boundary layers at all.
In addition, in the RBC cell, there may be reflections from the top and bottom plates, making it difficult to measure velocity fields in the vicinity.
The kinematic boundary layer thickness for $\Ra=10^7$ and $\Pr=\num{0.7}$ is estimated to be $\bl_\text{kin}\approx0.027$ \cite{shishkina_boundary_2010}.
Therefore, to further emulate a real PIV dataset, the data points within the kinematic boundary layer are omitted for training.
The resulting fields of velocity assimilation and temperature reconstruction on a subarea of plane $A$ are plotted in \cref{fig:plot_omitBL_fields}.
The left side shows the case where the PINN is trained with the spatially fully resolved DNS velocity field, and the right side shows the case, where the velocity data within the kinematic boundary layers is omitted.
The fields of the ground truth velocity component $v$ are plotted at the top, the reconstructed velocity fields are plotted in the middle, and the reconstructed temperature fields are plotted at the bottom.
Both reconstructed velocity fields (\cref{fig:plot_omitBL_fields}, center row) do not show differences in the reconstruction of the bulk flow region.
The PINN also manages to reconstruct the velocity field in the area where no data is given.
This is particularly evident in the reconstructed velocity field with omitted data labels (\cref{fig:plot_omitBL_fields}, center row, right), at $y\approx0.1$ near the bottom plate, where a rightward-streaming area is reconstructed despite its absence in the data.
The two reconstructed temperature fields also differ only slightly (\cref{fig:plot_omitBL_fields}, bottom row).
It can be seen that the PINN manages to reconstruct the thermal boundary layer structure despite the absence of data in this region ($\bl_\text{th}$ and $\bl_\text{kin}$ are close for $\Pr=0.7$); however the thermal boundary layer on the right side of \cref{fig:plot_omitBL_fields} seems to be thinner.

\begin{figure}[htbp]
 \centering
 \includegraphics{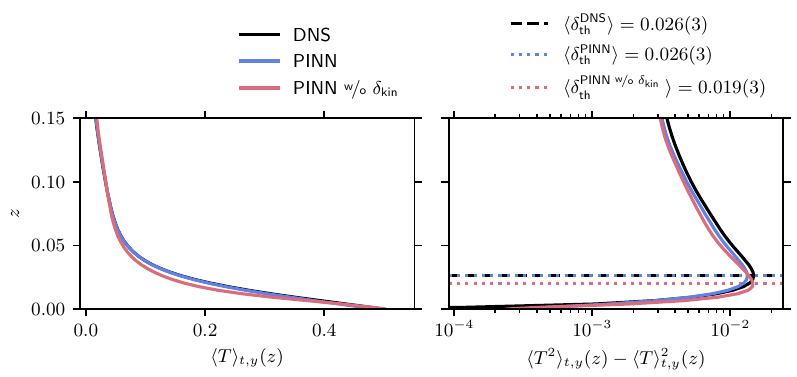}
 \caption{The temperature field on plane $A$, averaged over $t$ and $y$, and as a function of $z$ (left), along with its variance (right) over a total time span of $I_t=17.5$; comparison of the DNS temperature profile, the reconstructed profile and the reconstructed profile when the PINN is trained without access to the data inside the kinematic boundary layers.}
 \label{fig:plot_T_profile_omitBL}
\end{figure}
To examine whether the physical properties are preserved, the temperature profiles near the heating plate $\langle T\rangle_{t,y}(z)$, as a function of the height and averaged over a time span of $I_t = 17.5$ and $y$, are plotted in \cref{fig:plot_T_profile_omitBL} (left) for both cases and the DNS.
Indeed, we observe slight differences in the temperature profile (\cref{fig:plot_T_profile_omitBL}, left) and in the corresponding variance profile $\langle T^2\rangle_{t,y}(z)-\langle T\rangle_{t,y}^2(z)$ (\cref{fig:plot_T_profile_omitBL}, right), resulting in a thinner boundary layer thickness of $\bl_\text{th}^\text{PINN $\nicefrac{w}{o}$ $\bl_\text{kin}$}=\num{0.019\pm0.003}$ compared to $\bl_\text{th}^\text{PINN}=\bl_\text{th}^\text{DNS}=\num{0.026\pm0.003}$.

\begin{figure}[htbp]
 \centering
 \includegraphics{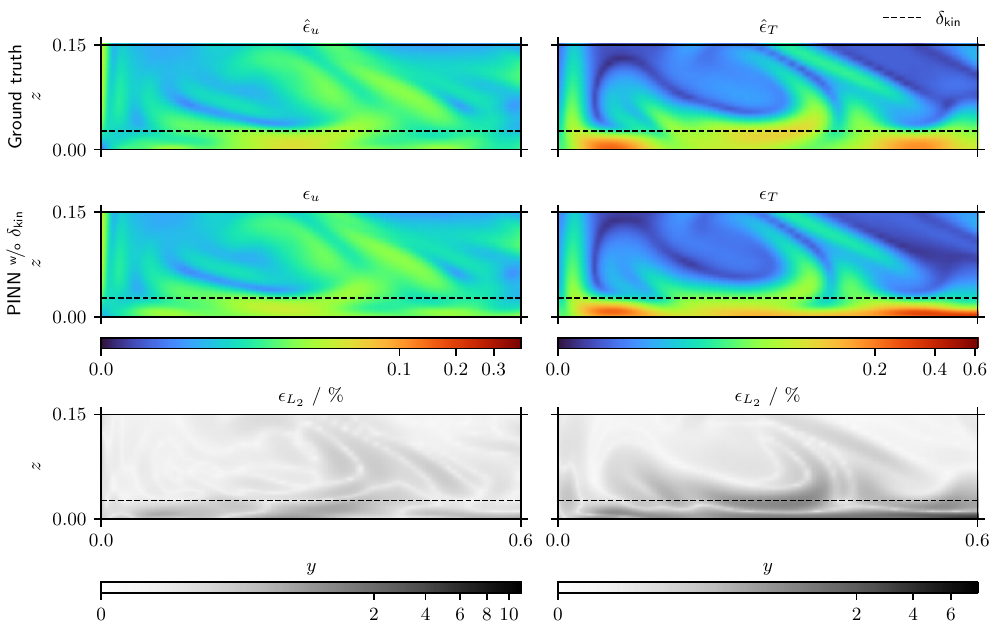}
 \caption{Viscous (left column) and thermal (right column) dissipation rates on a subset of the plane $A$. The ground truth fields ($\hat \cdot$) are shown at the top, the PINN predicted fields obtained by automatic differentiation are shown in the middle row, and the relative error fields are shown at the bottom. The PINN model is trained on velocity data, excluding kinematic boundary layer data.}
 \label{fig:plot_dissipation_woBL_zoom}
\end{figure}
Since the determination of dissipation rates is of great interest but experimentally challenging (see e.g. the experimental measurement of the viscous dissipation rate in \citeauthor{xu_experimental_2024} \cite{xu_experimental_2024}), we want to study the viscous ($\epsilon_u$, \cref{eq:viscous_diss}) and thermal ($\epsilon_T$, \cref{eq:thermal_diss}) dissipation rates, when reconstructed from the PINN model trained only on velocity data outside the kinematic boundary layers.
\Cref{fig:plot_dissipation_woBL_zoom} shows the ground truth dissipation fields ($\hat \cdot$, top row), the PINN reconstructed fields obtained by automatic differentiation (middle row), and the respective relative error fields (bottom row).
The kinematic dissipation rate is shown on the left and the thermal dissipation rate on the right.
Despite the lack of data in $\bl_\text{kin}$, the overall structures of both reconstructed dissipation rates are vely similar to the ground truth.
Compared to the PINN trained on velocity data in the whole plane (\cref{sec:dissipation}), we observe here the highest errors for both (viscous and thermal) dissipation rates inside the boundary layers.
While for the viscous dissipation the reconstructed fields inside the boundary layer (\cref{fig:plot_dissipation_woBL_zoom}, middle row, left) slightly lack intensity, the reconstructed thermal dissipation rate (\cref{fig:plot_dissipation_woBL_zoom}, middle row, right) shows higher values.
This is probably due to the imposed boundary conditions, which result in reduced velocities and increased temperature values.
The relative $\Ltwo$ errors computed with data on plane $A$ from the time instant shown in \cref{fig:plot_dissipation} are $\rLtwo(\epsilon_u)=\SI{41.9}{\percent}$ for the viscous dissipation and $\rLtwo(\epsilon_T)=\SI{68.3}{\percent}$ for the thermal dissipation.
To distinguish between the error contributions from the boundary layers and the bulk, we compute the relative $\Ltwo$ errors using only points that are either in the boundary layers or in the bulk.
For the viscous dissipation we get $\rLtwo(\epsilon_{u,\text{BL}})=\SI{50.5}{\percent}$ for the boundary layers and $\rLtwo(\epsilon_{u,\text{bulk}})=\SI{9.3}{\percent}$ for the bulk.
For the thermal dissipation we get $\rLtwo(\epsilon_{T,\text{BL}})=\SI{68.7}{\percent}$ and $\rLtwo(\epsilon_{T,\text{bulk}})=\SI{42.1}{\percent}$.
\begin{table}[htbp]
\centering
\begin{tabular}{@{}llSS@{}}
\toprule
                                     &         & \text{PINN} & \text{PINN $\nicefrac{\text{w}}{\text{o}}$ $\bl_\text{kin}$} \\ \midrule
\multirow{3}{*}{$\rLtwo(\epsilon_u)$ / \si{\percent}} & $A$ & 9.3 & 41.9                                                  \\
                                     & BL      & 9.2 & 50.5                                                  \\
                                     & bulk    & 9.5  & 9.3                                                   \\ \midrule
\multirow{3}{*}{$\rLtwo(\epsilon_T)$ / \si{\percent}} & $A$ & 19.7 & 68.3                                                  \\
                                     & BL      & 19.6 & 68.7                                                    \\
                                     & bulk    & 24.0  & 42.1                                                   \\ \bottomrule
\end{tabular}
\caption{Comparison of $\rLtwo$ errors when the PINN is trained with velocity data on plane $A$ and when it is trained with velocity data only outside the kinematic boundary layers ($\nicefrac{\text{w}}{\text{o}}$ $\bl_\text{kin}$). We distinguish between boundary layer (BL) and bulk contributions to the error.}
\label{tab:rLtwoDiss}
\end{table}
\Cref{tab:rLtwoDiss} shows the comparison of the $\rLtwo$ errors with the PINN trained on the velocity data in the whole plane (\cref{sec:dissipation}).
We observe that missing data within the kinematic boundary layers affects the accuracy of the dissipation rates in these regions.
However, the relative $\Ltwo$ error for viscous dissipation in the bulk remains unaffected, and the relative $\Ltwo$ error for thermal dissipation in the bulk is affected, but less than in the boundary layers.

\section{Conclusion and Outlook}
Assuming that 2D3C (two-dimensional, three-component) planar PIV velocity fields are available, we generate the missing temperature field using a PINN and DNS data as the ground truth.
We show, that training the PINN with three-component spatially (but not temporally) fully resolved velocity snapshots from DNS and the governing PDEs in an enclosing 3D layer yields promising results with the coefficient of determination exceeding $\Rtwo(T)>0.995$.
It turns out that careful considerations have to be done concerning the spatial training domain.
Having access to only planar data, the PINNs training domain in out-of-plane direction ($\delta_x$) needs to be larger than the size of the characteristic length of the medium to big eddies in the flow, i.e. that integral length scale, to be able to fully reconstruct the fields.
As the training domain becomes thinner, the agreement with the ground truth becomes increasingly poorer.
At $\delta_x=0$, some of the general structures in the temperature field are reconstructed, but it is not possible to obtain the correct out-of-plane derivatives $\nicefrac{\partial}{\partial x}$.
We apply a weighting of the PDE residuals by distributing collocation points in $x$-direction normally around the PIV plane, i.e. the region where data is provided for training is weighted highest.
Furthermore, we incorporate the knowledge of the DNS grid to define a sampling distribution in the vertical $z$-direction, placing greater emphasis on the boundary layer and its associated high temperature gradients.

A time span study reveals, that the results are worst at the beginning of the temporal training domain.
As initial conditions are not provided, the PINN requires a minimum time span exceeding an integral time scale of the flow to obtain reliable results.
This result is not only relevant for our specific case but also to other time-dependent assimilation and reconstruction tasks.
In our case, a dataset covering a total time span of half an LSC turnover is found to be a suitable choice.
In terms of temporal resolution, the PINN remains robust until the snapshot interval exceeds one half of the integral time scale.
This implies that the data must provide at least a framework in which all medium-scale time dynamics are resolved.
The PINN can then fill the gaps between the snapshots with physically reasonable fields that remain consistent with the given data.

We showed that the suggested approach allows to reconstruct large and small structures of the temperature as well as pressure fields.
We mimicked stereo PIV data by using only planar velocity fields.
In a common approach, we used Gaussian noise as a surrogate for potential measurement errors.
Even with a high noise ratio of $\sigma=\SI{30}{\percent}$, the PINN is still able to reconstruct the velocity with $\Rtwo(\{u,v,w\})>0.995$.
However, experimental data may not contain such an intense white noise.
Instead, it is subjected to colored noise or more systematic measurement uncertainties, such as zones with reflections and mirroring.
Since the areas at the lower and upper plate are mostly affected by such reflections and moreover only a few tracer particles reach the boundary layers, we also omitted the data labels inside the kinematic boundary layers.
It turns out that the PINN has no difficulties in reconstructing the fields if the data does not contain the top and bottom kinematic boundary layers.

In addition to reconstructing the temperature field, the PINN enables access to all partial derivatives of the physical fields through automatic differentiation.
This capability allows for the computation of viscous and thermal dissipation rates.
Evaluating the solution on plane $A$, we found that the PINN is fully able to reconstruct all visible structures in the viscous and thermal dissipation rates.
Even when trained without data inside the kinematic boundary layers, it still provides an estimate of the dissipation rates within these layers.
Moreover, the accuracy of viscous dissipation in the bulk region remains unaffected under these conditions.

Overall, we can state, that PINNs are fully able to reconstruct temperature fields of turbulent RBC, when having access to only planar snapshots of velocity data.
Not only temperature fields, but also viscous and thermal dissipation rates, can be obtained with the presented approach, opening up new possibilities for experimenters to study the physics of RBC through planar velocity measurements.

In our case, we used velocity data on a vertical plane; however, it would also be interesting to explore whether temperature reconstruction on a horizontal plane in the RBC cell could be achieved with the same level of accuracy.
The challenge here is likely to be greater due to the higher out-of-plane velocities, and therefore adjustments to the out-of-plane sampling would be required.
Another aspect worth investigating is whether temperature can be reconstructed using only mono-PIV data, i.e. the two in-plane velocity components only.
The PINN would likely require additional physical constraints, such as the net mass flux through the plane, to compensate for the lack of out-of-plane velocity information.
These aspects will be further investigated and developed in future work.

\appendix
\section{Layer and Neuron Variation Study}
\label{sec:layer_neuron_variation_study}
\begin{figure}[htbp]
 \centering
 \includegraphics{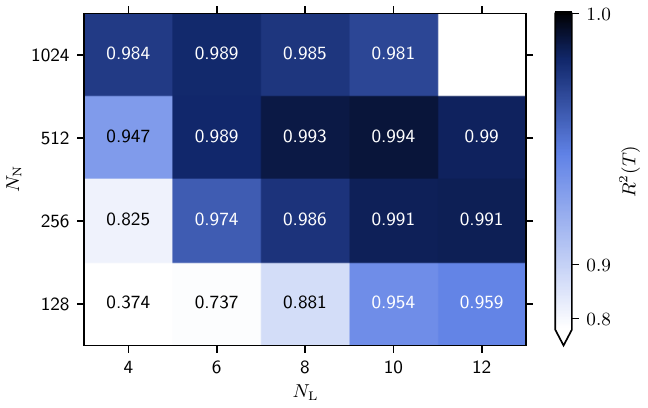}
 \caption{Results of the temperature reconstruction ($\Rtwo$) when varying the number of hidden layers ($N_\mathrm{L}$) and neurons per layer ($N_\mathrm{N}$).}
 \label{fig:plot_layer_neuron_matrix}
\end{figure}
\label{app:layer_neuron_variation}
Our default PINN configuration consists of $N_\mathrm{L}=10$ densely connected layers with $N_\mathrm{N}=256$ sine activated neurons each, which is the result of a parameter study.
\Cref{fig:plot_layer_neuron_matrix} shows the resulting $\Rtwo(T)$ values from PINNs with varying numbers of hidden layers and neurons per layer.
We observe a trend of increasing performance from the bottom left to the top right, corresponding to the use of more layers and neurons.
The smallest tested network of $N_\mathrm{L}=4$ layers and $N_\mathrm{N}=128$ neurons each finds the general pattern of the temperature and pressure fields, but has problems especially in regions with high gradients, leading to the lowest result of $\Rtwo(T)=0.374$.
The expressivity of the network is increased with higher numbers of layers and neurons, leading generally to better results in our case.
The best result of $\Rtwo(T)=\num{0.994}$ is obtained using $N_\mathrm{L}=10$ layers and $N_\mathrm{N}=512$ neurons.
Increasing the network width to more than $N_\mathrm{L}=10$ layers does not affect the results, and the same holds true for the number of neurons.
Good results are already achieved using $N_\mathrm{N}=256$ neurons.
Using more neurons is especially beneficial for smaller numbers of layers, but this effect vanishes for $N_\mathrm{L}>10$.
Increasing the number of neurons to $N_\mathrm{N}=1024$ leads to a significant increase in computational cost without leading to better $\Rtwo(T)$ values (cf. \cref{fig:plot_layer_neuron_matrix}).

\section{Training history}
\label{sec:training_history}
\begin{figure}[htbp]
 \includegraphics{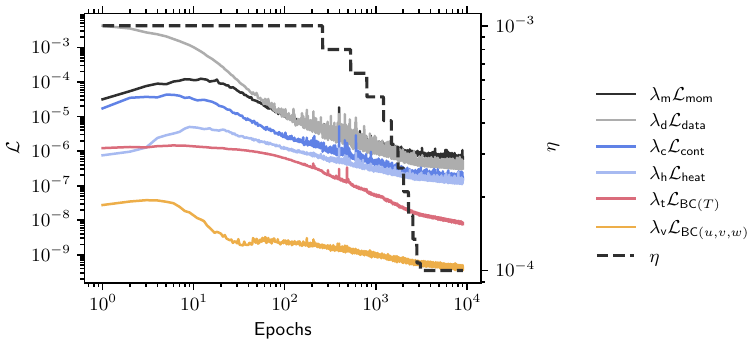}
 \caption{Evolution of the weighted training loss components and the learning rate ($\eta$) over epochs. For the weight values, refer to \cref{sec:optimization}.}
 \label{fig:plot_training_history_loss}
\end{figure}
\begin{figure}[htbp]
 \centering
 \includegraphics{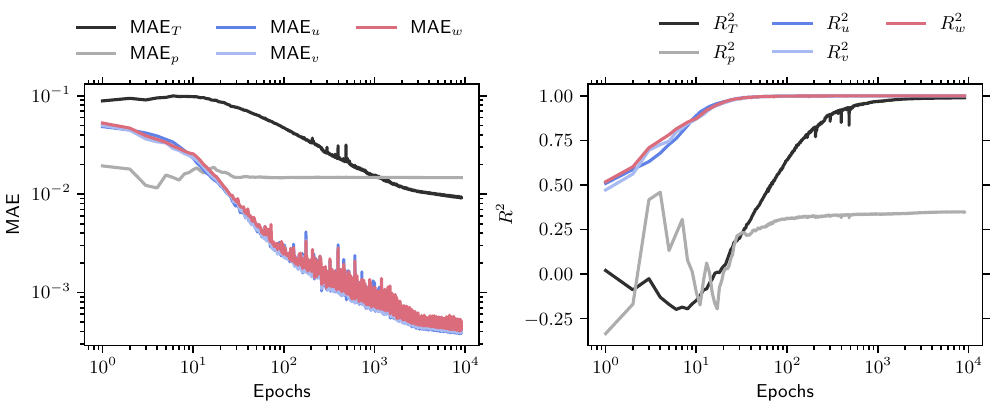}
 \caption{Evolution of the validation metrics $\MAE$ (left) and $\Rtwo$ (right) over epochs. \SI{5}{\percent} of the dataset is excluded from training and used for validation.}
 \label{fig:plot_training_history_metrics}
\end{figure}
An example of the evolution of the several weighted loss functions during training is shown in \cref{fig:plot_training_history_loss} and the corresponding tracking metrics $\MAE$ and $\Rtwo$ are plotted in \cref{fig:plot_training_history_metrics}.
The PINN is trained for $10^4$ epochs, starting with a learning rate of $\eta=\num{1e-3}$, which is gradually reduced to $\eta=\num{1e-4}$.
The loss weight values are given in \cref{sec:optimization}.
For the first ten epochs, only the data loss function $\mathcal{L}_\mathrm{d}$ decreases, whereas the PDE loss functions increase ($\mathcal{L}_\mathrm{mom}$, $\mathcal{L}_\mathrm{cont}$, $\mathcal{L}_\mathrm{energy}$) and the boundary condition loss components stagnate ($\mathcal{L}_{\mathrm{BC}(T)}$, $\mathcal{L}_{\mathrm{BC}(u,v,w)}$).
After this initial phase, the PINN begins optimizing the PDE and boundary condition losses.

The metric evaluation in \cref{fig:plot_training_history_metrics} confirms this observation: during the initial phase, the $\MAE$ metric for the velocity fields decreases (left), while the $\Rtwo$ values of the velocity field increase.
After approximately the first ten epochs, the metrics for the temperature field begin to improve.

The metrics for the pressure are generally the worst during the training, as the PDEs involve only the pressure gradient $\nabla p$ but not the absolute pressure values.

\section*{Acknowledgements}
The authors gratefully acknowledge the financial support of the program Procope-Mobilité of the French Embassy
in Germany and the scientific support and HPC resources provided by the German Aerospace Center (DLR).
The HPC system CARA is partially funded by “Saxon State Ministry for Economic Affairs, Labour and Transport“ and „Federal Ministry for Economic Affairs and Climate Action”.
The HPC system CARO is partially funded by “Ministry of Science and Culture of Lower Saxony“ and „Federal Ministry for Economic Affairs and Climate Action”.


\def\url#1{}  
\renewcommand{\harvardurl}{}
\bibliographystyle{abbrvnat}
\bibliography{biblio.bib}

\begin{thebibliography}{37}
\providecommand{\natexlab}[1]{#1}
\providecommand{\url}[1]{\texttt{#1}}
\expandafter\ifx\csname urlstyle\endcsname\relax
  \providecommand{\doi}[1]{doi: #1}\else
  \providecommand{\doi}{doi: \begingroup \urlstyle{rm}\Url}\fi

\bibitem[Dehne et~al.()Dehne, Lange, Schmeling, and
  Gores]{dehne_experimental_2024}
Tobias Dehne, Pascal Lange, Daniel Schmeling, and Ingo Gores.
\newblock Experimental evaluation of alternative ceiling-based ventilation
  systems for long-range passenger aircraft.
\newblock 15\penalty0 (4):\penalty0 1031--1050.
\newblock ISSN 1869-5590.
\newblock \doi{10.1007/s13272-024-00739-5}.
\newblock URL \url{https://doi.org/10.1007/s13272-024-00739-5}.

\bibitem[Schröder and Schanz()]{schroder_3d_2023}
Andreas Schröder and Daniel Schanz.
\newblock 3d lagrangian particle tracking in fluid mechanics.
\newblock 55:\penalty0 511--540.
\newblock ISSN 0066-4189, 1545-4479.
\newblock \doi{10.1146/annurev-fluid-031822-041721}.
\newblock URL
  \url{https://www.annualreviews.org/content/journals/10.1146/annurev-fluid-031822-041721}.
\newblock Publisher: Annual Reviews.

\bibitem[Barta et~al.()Barta, Bauer, Herzog, Schiepel, and
  Wagner]{barta_proptv_2024}
Robin Barta, Christian Bauer, Sebastian Herzog, Daniel Schiepel, and Claus
  Wagner.
\newblock {proPTV}: A probability-based particle tracking velocimetry
  framework.
\newblock 514:\penalty0 113212.
\newblock ISSN 0021-9991.
\newblock \doi{10.1016/j.jcp.2024.113212}.
\newblock URL
  \url{https://www.sciencedirect.com/science/article/pii/S0021999124004613}.

\bibitem[Schiepel et~al.({\natexlab{a}})Schiepel, Herzog, and
  Wagner]{schiepel_experimental_2018}
Daniel Schiepel, Sebastian Herzog, and Claus Wagner.
\newblock Experimental study of turbulent rayleigh-bénard convection using
  large-scale tomo-{PIV} and high-density {PTV}.
\newblock In Michel~O. Deville, Vincent Couaillier, Jean-Luc Estivalezes,
  Vincent Gleize, Thiên {HiêpLê}, Marc Terracol, and Stéphane Vincent,
  editors, \emph{Turbulence and Interactions}, pages 225--231. Springer
  International Publishing, {\natexlab{a}}.
\newblock ISBN 978-3-319-60387-2.
\newblock \doi{10.1007/978-3-319-60387-2_24}.

\bibitem[Schmeling et~al.({\natexlab{a}})Schmeling, Czapp, Bosbach, and
  Wagner]{schmeling_development_2010}
Daniel Schmeling, Marek Czapp, Johannes Bosbach, and Claus Wagner.
\newblock Development of combined particle image velocimetry and particle image
  thermography for air flows.
\newblock In \emph{2010 14th International Heat Transfer Conference}, pages
  57--64, {\natexlab{a}}.
\newblock ISBN 978-0-7918-3879-2.
\newblock URL \url{http://www.asmeconferences.org/ihtc14/}.
\newblock Issue: {IHTC}14-22774 Number: {IHTC}14-22774.

\bibitem[Schiepel et~al.({\natexlab{b}})Schiepel, Schmeling, and
  Wagner]{schiepel_simultaneous_2021}
Daniel Schiepel, Daniel Schmeling, and Claus Wagner.
\newblock Simultaneous tomographic particle image velocimetry and thermometry
  of turbulent rayleigh–bénard convection.
\newblock 32\penalty0 (9):\penalty0 095201, {\natexlab{b}}.
\newblock ISSN 0957-0233.
\newblock \doi{10.1088/1361-6501/abf095}.
\newblock URL \url{https://dx.doi.org/10.1088/1361-6501/abf095}.
\newblock Publisher: {IOP} Publishing.

\bibitem[Käufer and Cierpka()]{kaufer_volumetric_2023}
T~Käufer and C~Cierpka.
\newblock Volumetric lagrangian temperature and velocity measurements with
  thermochromic liquid crystals.
\newblock 35\penalty0 (3):\penalty0 035301.
\newblock ISSN 0957-0233.
\newblock \doi{10.1088/1361-6501/ad16d1}.
\newblock URL \url{https://dx.doi.org/10.1088/1361-6501/ad16d1}.
\newblock Publisher: {IOP} Publishing.

\bibitem[Schmeling et~al.({\natexlab{b}})Schmeling, Bosbach, and
  Wagner]{schmeling_simultaneous_2014}
Daniel Schmeling, Johannes Bosbach, and Claus Wagner.
\newblock Simultaneous measurement of temperature and velocity fields in
  convective air flows.
\newblock 25\penalty0 (3):\penalty0 035302, {\natexlab{b}}.
\newblock ISSN 0957-0233.
\newblock \doi{10.1088/0957-0233/25/3/035302}.
\newblock URL \url{https://dx.doi.org/10.1088/0957-0233/25/3/035302}.
\newblock Publisher: {IOP} Publishing.

\bibitem[Gasteuil et~al.()Gasteuil, Shew, Gibert, Chillá, Castaing, and
  Pinton]{gasteuil_lagrangian_2007}
Y.~Gasteuil, W.~L. Shew, M.~Gibert, F.~Chillá, B.~Castaing, and J.-F. Pinton.
\newblock Lagrangian temperature, velocity, and local heat flux measurement in
  rayleigh-bénard convection.
\newblock 99\penalty0 (23):\penalty0 234302.
\newblock \doi{10.1103/PhysRevLett.99.234302}.
\newblock URL \url{https://link.aps.org/doi/10.1103/PhysRevLett.99.234302}.
\newblock Publisher: American Physical Society.

\bibitem[Clark Di~Leoni et~al.()Clark Di~Leoni, Mazzino, and
  Biferale]{clark_di_leoni_synchronization_2020}
Patricio Clark Di~Leoni, Andrea Mazzino, and Luca Biferale.
\newblock Synchronization to big data: Nudging the navier-stokes equations for
  data assimilation of turbulent flows.
\newblock 10\penalty0 (1):\penalty0 011023.
\newblock \doi{10.1103/PhysRevX.10.011023}.
\newblock URL \url{https://link.aps.org/doi/10.1103/PhysRevX.10.011023}.
\newblock Publisher: American Physical Society.

\bibitem[Bauer et~al.()Bauer, Schiepel, and Wagner]{bauer_assimilation_2022}
C.~Bauer, D.~Schiepel, and C.~Wagner.
\newblock Assimilation and extension of particle image velocimetry data of
  turbulent rayleigh–bénard convection using direct numerical simulations.
\newblock 63\penalty0 (1):\penalty0 22.
\newblock ISSN 1432-1114.
\newblock \doi{10.1007/s00348-021-03369-3}.
\newblock URL \url{https://doi.org/10.1007/s00348-021-03369-3}.

\bibitem[Weiss et~al.()Weiss, Emran, Bosbach, and
  Shishkina]{weiss_temperature_2025}
Stephan Weiss, Mohammad~S. Emran, Johannes Bosbach, and Olga Shishkina.
\newblock On temperature reconstruction from velocity fields in turbulent
  rayleigh–bénard convection.
\newblock 242:\penalty0 126768.
\newblock ISSN 0017-9310.
\newblock \doi{10.1016/j.ijheatmasstransfer.2025.126768}.
\newblock URL
  \url{https://www.sciencedirect.com/science/article/pii/S0017931025001097}.

\bibitem[Raissi et~al.()Raissi, Perdikaris, and
  Karniadakis]{raissi_physics-informed_2019}
M.~Raissi, P.~Perdikaris, and G.~E. Karniadakis.
\newblock Physics-informed neural networks: A deep learning framework for
  solving forward and inverse problems involving nonlinear partial differential
  equations.
\newblock 378:\penalty0 686--707.
\newblock ISSN 0021-9991.
\newblock \doi{10.1016/j.jcp.2018.10.045}.
\newblock URL
  \url{https://www.sciencedirect.com/science/article/pii/S0021999118307125}.

\bibitem[Wassing et~al.()Wassing, Langer, and
  Bekemeyer]{wassing_physics-informed_2024}
Simon Wassing, Stefan Langer, and Philipp Bekemeyer.
\newblock Physics-informed neural networks for parametric compressible euler
  equations.
\newblock 270:\penalty0 106164.
\newblock ISSN 0045-7930.
\newblock \doi{10.1016/j.compfluid.2023.106164}.
\newblock URL
  \url{https://www.sciencedirect.com/science/article/pii/S0045793023003894}.

\bibitem[Cai et~al.()Cai, Gray, and Karniadakis]{cai_physics-informed_2024}
Shengze Cai, Callum Gray, and George~Em Karniadakis.
\newblock Physics-informed neural networks enhanced particle tracking
  velocimetry: An example for turbulent jet flow.
\newblock 73:\penalty0 1--9.
\newblock ISSN 1557-9662.
\newblock \doi{10.1109/TIM.2024.3398068}.
\newblock URL \url{https://ieeexplore.ieee.org/abstract/document/10522764}.
\newblock Conference Name: {IEEE} Transactions on Instrumentation and
  Measurement.

\bibitem[Wang et~al.()Wang, Liu, and Wang]{wang_dense_2022}
Hongping Wang, Yi~Liu, and Shizhao Wang.
\newblock Dense velocity reconstruction from particle image
  velocimetry/particle tracking velocimetry using a physics-informed neural
  network.
\newblock 34\penalty0 (1):\penalty0 017116.
\newblock ISSN 1070-6631.
\newblock \doi{10.1063/5.0078143}.
\newblock URL \url{https://doi.org/10.1063/5.0078143}.

\bibitem[Steinfurth et~al.()Steinfurth, Hassanein, Doan, and
  Scarano]{steinfurth_physics-informed_2024}
B.~Steinfurth, A.~Hassanein, N.~A.~K. Doan, and F.~Scarano.
\newblock Physics-informed neural networks for dense reconstruction of vortex
  rings from particle tracking velocimetry.
\newblock 36\penalty0 (9):\penalty0 095110.
\newblock ISSN 1070-6631.
\newblock \doi{10.1063/5.0212585}.
\newblock URL \url{https://doi.org/10.1063/5.0212585}.

\bibitem[Sundar et~al.()Sundar, Majumdar, Lucor, and
  Sarkar]{sundar_physics-informed_2024}
Rahul Sundar, Dipanjan Majumdar, Didier Lucor, and Sunetra Sarkar.
\newblock Physics-informed neural networks modelling for systems with moving
  immersed boundaries: Application to an unsteady flow past a plunging foil.
\newblock 125:\penalty0 104066.
\newblock ISSN 0889-9746.
\newblock \doi{10.1016/j.jfluidstructs.2024.104066}.
\newblock URL
  \url{https://www.sciencedirect.com/science/article/pii/S088997462400001X}.

\bibitem[Toscano et~al.()Toscano, Käufer, Wang, Maxey, Cierpka, and
  Karniadakis]{toscano_inferring_2024}
Juan~Diego Toscano, Theo Käufer, Zhibo Wang, Martin Maxey, Christian Cierpka,
  and George~Em Karniadakis.
\newblock Inferring turbulent velocity and temperature fields and their
  statistics from lagrangian velocity measurements using physics-informed
  kolmogorov-arnold networks.
\newblock URL \url{http://arxiv.org/abs/2407.15727}.

\bibitem[Mommert et~al.()Mommert, Barta, Bauer, Volk, and
  Wagner]{mommert_periodically_2024}
Michael Mommert, Robin Barta, Christian Bauer, Marie-Christine Volk, and Claus
  Wagner.
\newblock Periodically activated physics-informed neural networks for
  assimilation tasks for three-dimensional rayleigh–bénard convection.
\newblock 283:\penalty0 106419.
\newblock ISSN 0045-7930.
\newblock \doi{10.1016/j.compfluid.2024.106419}.
\newblock URL
  \url{https://www.sciencedirect.com/science/article/pii/S0045793024002500}.

\bibitem[Lucor et~al.()Lucor, Agrawal, and Sergent]{lucor_simple_2022}
Didier Lucor, Atul Agrawal, and Anne Sergent.
\newblock Simple computational strategies for more effective physics-informed
  neural networks modeling of turbulent natural convection.
\newblock 456:\penalty0 111022.
\newblock ISSN 0021-9991.
\newblock \doi{10.1016/j.jcp.2022.111022}.
\newblock URL
  \url{https://www.sciencedirect.com/science/article/pii/S0021999122000845}.

\bibitem[Feldmann et~al.()Feldmann, , and Wagner]{feldmann_direct_2012}
Daniel Feldmann, , and Claus Wagner.
\newblock Direct numerical simulation of fully developed turbulent and
  oscillatory pipe flows at.
\newblock 13:\penalty0 N32.
\newblock ISSN null.
\newblock \doi{10.1080/14685248.2012.708470}.
\newblock URL \url{https://doi.org/10.1080/14685248.2012.708470}.
\newblock Publisher: Taylor \& Francis.

\bibitem[Shishkina et~al.()Shishkina, Stevens, Grossmann, and
  Lohse]{shishkina_boundary_2010}
Olga Shishkina, Richard J. A.~M. Stevens, Siegfried Grossmann, and Detlef
  Lohse.
\newblock Boundary layer structure in turbulent thermal convection and its
  consequences for the required numerical resolution.
\newblock 12\penalty0 (7):\penalty0 075022.
\newblock ISSN 1367-2630.
\newblock \doi{10.1088/1367-2630/12/7/075022}.
\newblock URL \url{https://dx.doi.org/10.1088/1367-2630/12/7/075022}.

\bibitem[Baydin et~al.()Baydin, Pearlmutter, Radul, and
  Siskind]{baydin_automatic_2018}
Atilim~Gunes Baydin, Barak~A. Pearlmutter, Alexey~Andreyevich Radul, and
  Jeffrey~Mark Siskind.
\newblock Automatic differentiation in machine learning: a survey.
\newblock 18\penalty0 (153):\penalty0 1--43.
\newblock ISSN 1533-7928.
\newblock URL \url{http://jmlr.org/papers/v18/17-468.html}.

\bibitem[Sitzmann et~al.()Sitzmann, Martel, Bergman, Lindell, and
  Wetzstein]{sitzmann_implicit_2020}
Vincent Sitzmann, Julien N.~P. Martel, Alexander~W. Bergman, David~B. Lindell,
  and Gordon Wetzstein.
\newblock Implicit neural representations with periodic activation functions.
\newblock URL \url{http://arxiv.org/abs/2006.09661}.

\bibitem[Nguyen et~al.()Nguyen, Dairay, Meunier, Millet, and
  Mougeot]{nguyen_fixed-budget_2023}
Thi Nguyen~Khoa Nguyen, Thibault Dairay, Raphaël Meunier, Christophe Millet,
  and Mathilde Mougeot.
\newblock Fixed-budget online adaptive learning for physics-informed neural
  networks. towards parameterized problem inference.
\newblock In Jiří Mikyška, Clélia de~Mulatier, Maciej Paszynski, Valeria~V.
  Krzhizhanovskaya, Jack~J. Dongarra, and Peter~M.A. Sloot, editors,
  \emph{Computational Science – {ICCS} 2023}, pages 453--468. Springer Nature
  Switzerland.
\newblock ISBN 978-3-031-36027-5.
\newblock \doi{10.1007/978-3-031-36027-5_36}.

\bibitem[Wu et~al.()Wu, Zhu, Tan, Kartha, and Lu]{wu_comprehensive_2023}
Chenxi Wu, Min Zhu, Qinyang Tan, Yadhu Kartha, and Lu~Lu.
\newblock A comprehensive study of non-adaptive and residual-based adaptive
  sampling for physics-informed neural networks.
\newblock 403:\penalty0 115671.
\newblock ISSN 0045-7825.
\newblock \doi{10.1016/j.cma.2022.115671}.
\newblock URL
  \url{https://www.sciencedirect.com/science/article/pii/S0045782522006260}.

\bibitem[Parzen()]{parzen_estimation_1962}
Emanuel Parzen.
\newblock On estimation of a probability density function and mode.
\newblock 33\penalty0 (3):\penalty0 1065--1076.
\newblock ISSN 0003-4851.
\newblock URL \url{https://www.jstor.org/stable/2237880}.
\newblock Publisher: Institute of Mathematical Statistics.

\bibitem[Stein()]{stein_large_1987}
M.~Stein.
\newblock Large sample properties of simulations using latin hypercube
  sampling.
\newblock 29\penalty0 (2):\penalty0 143--151.
\newblock \doi{10.1080/00401706.1987.10488205}.

\bibitem[Kingma and Ba()]{kingma_adam_2017}
Diederik~P. Kingma and Jimmy Ba.
\newblock Adam: A method for stochastic optimization.
\newblock URL \url{http://arxiv.org/abs/1412.6980}.

\bibitem[Thacker et~al.()Thacker, Loyer, and Aubrun]{thacker_comparison_2010}
A.~Thacker, S.~Loyer, and S.~Aubrun.
\newblock Comparison of turbulence length scales assessed with three
  measurement systems in increasingly complex turbulent flows.
\newblock 34\penalty0 (5):\penalty0 638--645.
\newblock ISSN 0894-1777.
\newblock \doi{10.1016/j.expthermflusci.2009.12.005}.
\newblock URL
  \url{https://www.sciencedirect.com/science/article/pii/S0894177709002076}.

\bibitem[Shannon()]{shannon_communication_1949}
C.E. Shannon.
\newblock Communication in the presence of noise.
\newblock 37\penalty0 (1):\penalty0 10--21.
\newblock ISSN 2162-6634.
\newblock \doi{10.1109/JRPROC.1949.232969}.
\newblock URL \url{https://ieeexplore.ieee.org/document/1697831}.
\newblock Conference Name: Proceedings of the {IRE}.

\bibitem[Kaczorowski et~al.()Kaczorowski, Chong, and
  Xia]{kaczorowski_turbulent_2014}
Matthias Kaczorowski, Kai-Leong Chong, and Ke-Qing Xia.
\newblock Turbulent flow in the bulk of rayleigh–bénard convection:
  aspect-ratio dependence of the small-scale properties.
\newblock 747:\penalty0 73--102.
\newblock ISSN 0022-1120, 1469-7645.
\newblock \doi{10.1017/jfm.2014.154}.
\newblock URL
  \url{https://www.cambridge.org/core/journals/journal-of-fluid-mechanics/article/turbulent-flow-in-the-bulk-of-rayleighbenard-convection-aspectratio-dependence-of-the-smallscale-properties/B9489FD3807D17048C999C12FBDA54CB}.

\bibitem[Shraiman and Siggia()]{shraiman_heat_1990}
Boris~I. Shraiman and Eric~D. Siggia.
\newblock Heat transport in high-rayleigh-number convection.
\newblock 42\penalty0 (6):\penalty0 3650--3653.
\newblock \doi{10.1103/PhysRevA.42.3650}.
\newblock URL \url{https://link.aps.org/doi/10.1103/PhysRevA.42.3650}.
\newblock Publisher: American Physical Society.

\bibitem[Grossmann and Lohse()]{grossmann_scaling_2000}
Siegfried Grossmann and Detlef Lohse.
\newblock Scaling in thermal convection: a unifying theory.
\newblock 407:\penalty0 27--56.
\newblock ISSN 1469-7645, 0022-1120.
\newblock \doi{10.1017/S0022112099007545}.
\newblock URL
  \url{https://www.cambridge.org/core/journals/journal-of-fluid-mechanics/article/scaling-in-thermal-convection-a-unifying-theory/C04F99EF099F794FC23B4939CCDB477F}.
\newblock Publisher: Cambridge University Press.

\bibitem[Stevens et~al.()Stevens, Poel, Grossmann, and
  Lohse]{stevens_unifying_2013}
Richard J. A.~M. Stevens, Erwin P. van~der Poel, Siegfried Grossmann, and
  Detlef Lohse.
\newblock The unifying theory of scaling in thermal convection: the updated
  prefactors.
\newblock 730:\penalty0 295--308.
\newblock ISSN 0022-1120, 1469-7645.
\newblock \doi{10.1017/jfm.2013.298}.
\newblock URL
  \url{https://www.cambridge.org/core/journals/journal-of-fluid-mechanics/article/unifying-theory-of-scaling-in-thermal-convection-the-updated-prefactors/71CC88EE08E81AA678985F5CCC1F45A2}.

\bibitem[Xu et~al.()Xu, Zhang, and Xia]{xu_experimental_2024}
Fang Xu, Lu~Zhang, and Ke-Qing Xia.
\newblock Experimental measurement of spatio-temporally resolved energy
  dissipation rate in turbulent rayleigh–bénard convection.
\newblock 984:\penalty0 A8.
\newblock ISSN 0022-1120, 1469-7645.
\newblock \doi{10.1017/jfm.2024.164}.
\newblock URL
  \url{https://www.cambridge.org/core/journals/journal-of-fluid-mechanics/article/experimental-measurement-of-spatiotemporally-resolved-energy-dissipation-rate-in-turbulent-rayleighbenard-convection/359C262A9092111FCCE5B786C1F3CCA9}.

\end{thebibliography}

\end{document}